\documentclass{nature}

\usepackage{caption}
\usepackage{graphicx}
\usepackage[scaled]{helvet}
\usepackage{amsmath}
\usepackage{amssymb}
\usepackage{bm}
\usepackage{xfrac}
\newcommand*{\myfont}{\fontfamily{phv}\selectfont}
\usepackage{hyperref}
\usepackage{anyfontsize}
\usepackage{ragged2e}

\usepackage{lineno}
 \usepackage{color}

\definecolor{CLBlue}{rgb}{0, .25, .8}
\definecolor{MyBlue}{rgb}{0, .24, .40}
\definecolor{MyTurquoise}{rgb}{0, .53, .49}
\definecolor{MyGreen}{rgb}{0, .35, 0}
\definecolor{MyOrange}{rgb}{.8, .46, 0}
\definecolor{MyRed}{rgb}{.57, .07, 0}
\definecolor{MyPurple}{rgb}{.46, .1, .46}

\title{Direct dependencies between neurons explain activity}

\author{Christopher W. Lynn$^{1,2,3}$}

\begin{document}


\maketitle

\begin{affiliations}
\item Department of Physics, Yale University, New Haven, CT 06520, USA
\item Quantitative Biology Institute, Yale University, New Haven, CT 06520, USA
\item Wu Tsai Institute, Yale University, New Haven, CT 06520, USA
\end{affiliations}


\noindent {\large \myfont \textbf{Abstract}}
\vspace{-28pt}

\noindent\rule{\textwidth}{.5pt}

\noindent Our understanding of neural computation is founded on the assumption that neurons fire in response to a linear summation of inputs. Yet experiments demonstrate that some neurons are capable of complex functions that require interactions between inputs. Here we show, across multiple brain regions and species, that direct dependencies (without interactions between inputs) explain most of the variability in neuronal activity. Neurons are quantitatively described by models that capture the measured dependence on each input individually, but assume nothing about combinations of inputs. These minimal models, which are equivalent to logistic artificial neurons, predict complex higher-order dependencies and recover known features of synaptic connectivity. The inferred neural network is sparse, indicating a highly redundant neural code that is robust to perturbations. These results suggest that, despite intricate biophysical details, most neurons are described by simple artificial models.



\newpage

\noindent {\large \myfont \textbf{Main}}
\vspace{-28pt}

\noindent\rule{\textwidth}{.5pt}

Neurons receive synaptic inputs from tens of other neurons in the roundworm \textit{C. elegans},\cite{White-01} hundreds to thousands in the fruit fly,\cite{Lin-02} and thousands to tens of thousands in mice, monkeys, and humans.\cite{Loomba-01} As the number of inputs grows, the space of possible computations explodes exponentially. To tame this complexity, simplified models have long assumed that the output of a neuron depends directly on individual inputs, without interactions between inputs.\cite{Mcculloch-01, Hopfield-01, Rosenblatt-01, Hertz-01, Hopfield-02, Amit-01} While this picture forms the foundation for our understanding of neural computation, both in the brain and artificial networks, it remains unclear whether direct dependencies alone can explain the activity of real neurons.


Across different species and neural systems, it is now possible to make long, stable recordings of large contiguous populations of neurons.\cite{Stringer-01, Gauthier-01, Dag-01, Demas-01, Urai-01} This means that for each neuron, we may have access to its output and all of its inputs simultaneously. But how can we determine whether the activity of each neuron arises from simple dependencies on individual inputs, or instead, requires complex interactions between inputs? To answer this question, we need a framework to quantify the minimal consequences of direct dependencies. \\


\noindent {\myfont \large Minimal consequences of direct dependencies}

\begin{figure}[t!]
\centering
\includegraphics[width = .8\textwidth]{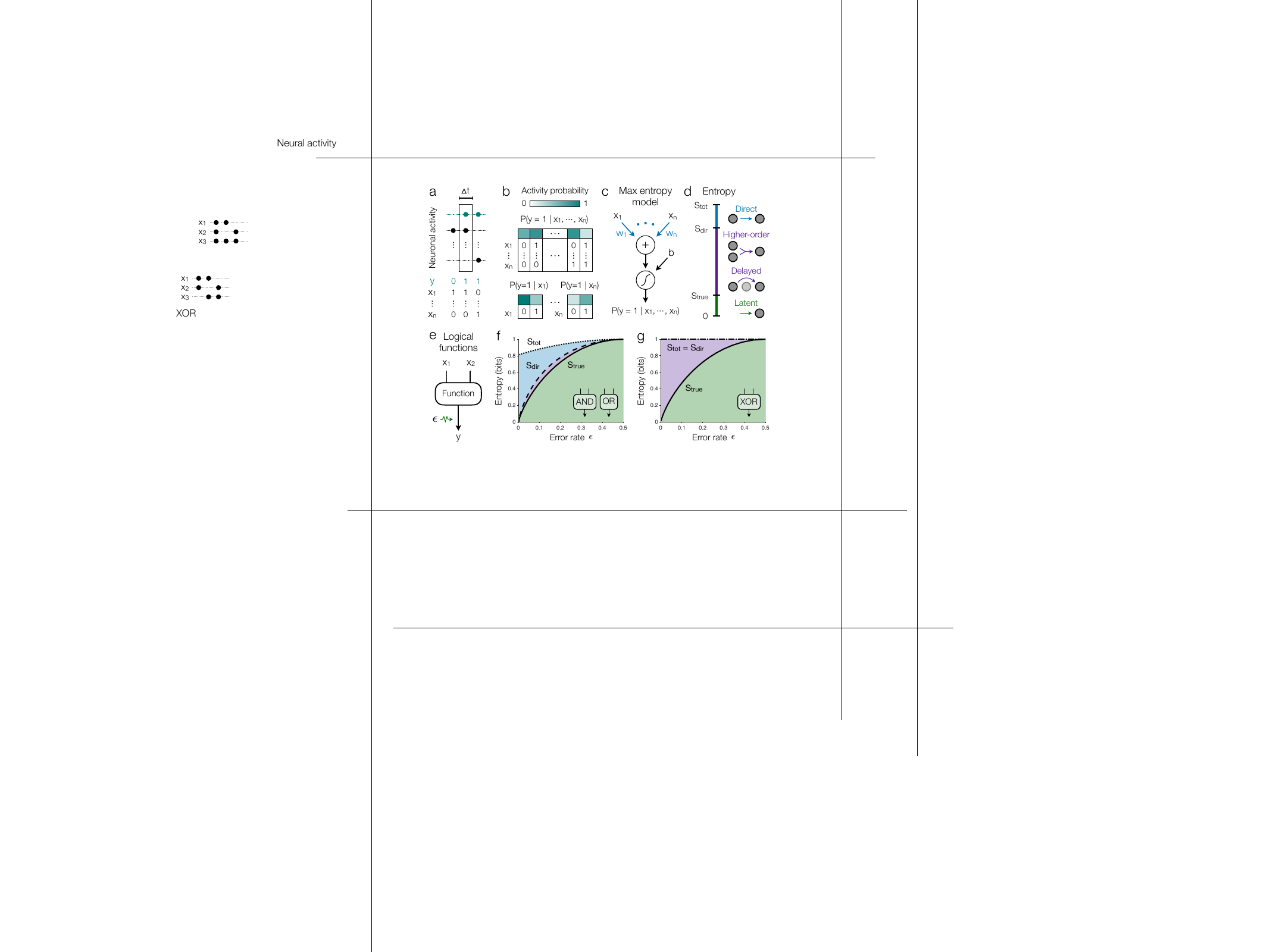} \\
\raggedright
\captionsetup{labelformat=empty}
{\spacing{1.25} \caption{\small \textbf{Fig.~\ref{fig_maxEnt} $|$ Quantifying the variability explained by direct dependencies.} \textbf{a}, Activity (dots) of an output neuron $y$ and $n$ inputs $x_1,\hdots,x_n$. Within a window of width $\Delta t$, each neuron binarizes into active ($y=1$) or silent ($y = 0$).\cite{Rieke-01, Schneidman-01} To study activity on the fastest available timescale, $\Delta t$ is defined by the sampling interval of a given experiment. \textbf{b}, The full input-output dependence is defined by the probability of activity in response to each of the $2^n$ combinations of inputs (\textit{top}). The simplest dependencies reflect the direct responses to each input individually, the number of which grows linearly with $n$ (\textit{bottom}). \textbf{c}, The minimal model, which has maximum entropy consistent with these direct dependencies,\cite{Jaynes-01, Cover-01} is equivalent to a logistic artificial neuron.\cite{Hertz-01} \textbf{d}, Hierarchy of entropies, where the difference $S_\text{tot} - S_\text{dir}$ quantifies the amount of variability captured by direct dependencies, $S_\text{dir} - S_\text{true}$ the variability due to higher-order and time-delayed dependencies, and $S_\text{true}$ the latent variability that cannot be explained by the inputs.\cite{Schneidman-02} \textbf{e}, Logical function with error probability $\epsilon$ and binary inputs that are drawn independently at random. \textbf{f-g}, Hierarchy of entropies versus error rate for the AND and OR functions (\textbf{f}) and the XOR function (\textbf{g}). For AND and OR (\textbf{f}), the functions are almost exactly captured by direct dependencies ($S_\text{dir} \approx S_\text{true}$), while for XOR (\textbf{g}), direct dependencies explain none of the output variability ($S_\text{dir} = S_\text{tot}$). \label{fig_maxEnt}}}
\end{figure}

Neurons in the brain receive inputs at their dendrites and then execute a binary function: they either remain silent ($y = 0$) or generate a discrete impulse ($y = 1$) known as an action potential or spike (Fig.~\ref{fig_maxEnt}a).\cite{Rieke-01, Schneidman-01} This mapping from inputs to output involves intricate details of membrane potential dynamics and cell morphology, which vary between neurons, brain regions, and species.\cite{Jan-01, Poirazi-01, Petersen-01} Yet for every neuron, these details culminate in a table of firing probabilities $P(y=1|x_1,\hdots,x_n)$, which define the function that the cell performs on its $n$ inputs $x_1,\hdots,x_n$. This description, while general enough to capture every neuron, is also hopelessly complex; it requires specifying a different firing probability for each combination of inputs, a number that grows exponentially with $n$ (Fig.~\ref{fig_maxEnt}b). To understand the functions performed by real neurons, we thus need simplifying hypotheses.

The simplest relationships between the output $y$ and inputs $x_i$ are contained in the direct dependencies $P(y|x_i)$; these capture everything about the activity that does not involve interactions between inputs. But how can we tell whether these simple dependencies are enough to describe a real neuron? Even if we measure all the direct dependencies $P(y|x_1),\hdots,P(y|x_n)$, there are still an infinite number of possible functions $P(y|x_1,\hdots,x_n)$ consistent with these constraints. Our problem, therefore, is to find the model that matches these simple dependencies, but is maximally random with regard to higher-order dependencies involving two, three, or more inputs. We show that this minimal model is the one with maximum entropy consistent with the average activity of $y$ and its correlations with the inputs $x_i$ (Methods).\cite{Jaynes-01, Cover-01} This maximum entropy model is known to take the form,
\begin{equation}
\label{eq_P}
P(y=1|x_1,\hdots,x_n) = \sigma\Big(b + \sum_{i=1}^n w_ix_i \Big),
\end{equation}
where $\sigma(\cdot)$ is the logistic function.\cite{Berger-01, Huang-01} The parameters $b$ and $w_i$ must be computed so that the model matches the measured direct dependencies, and we provide an algorithm that converges efficiently, even for large $n$ (Methods).

Logistic models (and other generalized linear models) have been used extensively to study the statistical dependencies of neural activity on other neurons as well as latent variables like stimuli, behavior, and arousal.\cite{Pillow-01, Goris-01, Stevenson-01, Weber-01} In fact, these models have been shown to capture key phenomenological features of neuronal spiking and firing rates.\cite{Ostojic-01, Mensi-01, Churchland-01, Priebe-01} The maximum entropy principle adds to this context a concrete mathematical connection between biological and artificial neurons: a real neuron with purely direct dependencies is equivalent to an artificial neuron with bias $b$, linear weights $w_i$, and logistic activation function (Fig.~\ref{fig_maxEnt}c).\cite{Hertz-01} Consequently, all other models---for example, with different activation functions, dependencies on latent variables, temporal dependencies, or regularized parameters---must involve sources of order beyond direct dependencies.


We are now prepared to study the amount of variability captured by direct dependencies. The total variability of a neuron is quantified by the entropy $S_\text{tot}$ of its output with no knowledge of the inputs.\cite{Strong-01} By contrast, the true entropy of a neuron $S_\text{true}$ quantifies the latent variability in its activity that cannot be explained by the inputs (Methods).\cite{Cover-01} With knowledge of only the direct dependencies, the entropy $S_\text{dir}$ of the minimal model in Eq.~(\ref{eq_P}) sits between these two extremes (Methods), yielding a hierarchy $S_\text{tot} \ge S_\text{dir} \ge S_\text{true} \ge 0$. In this way, the difference $S_\text{tot} - S_\text{dir}$ defines the amount of variability captured only by direct dependencies. The remaining variability $S_\text{dir}$ can arise from any other dependence on the inputs (including higher-order and time-delayed dependencies) or latent variables that are not included in the inputs (Fig.~\ref{fig_maxEnt}d).\cite{Schneidman-02} Importantly, if the direct entropy $S_\text{dir}$ becomes small, then so too does the true entropy $S_\text{true}$. In this limit, the model (and thus the neuron itself) becomes equivalent to a McCulloch-Pitts (MP) neuron,\cite{Mcculloch-01} or a perceptron,\cite{Block-01} and all the variability is explained by direct dependencies (Methods).

To gain intuition, consider an output $y$ that performs a logical function on two binary inputs $x_1$ and $x_2$ with error rate $\epsilon$ (Fig.~\ref{fig_maxEnt}e). For AND and OR gates, as $\epsilon$ increases, the output becomes more stochastic, leading to higher variability (Fig.~\ref{fig_maxEnt}f). Across all error rates $\epsilon$, we find that the direct entropy $S_\text{dir}$ lies close to the true entropy $S_\text{true}$, such that the model provides a tight approximation to the true computation. As errors vanish and the output becomes deterministic ($\epsilon = 0$), the minimal model becomes exact ($S_\text{dir} = S_\text{true}$) and all the variability in $y$ is explained by direct dependencies ($S_\text{dir} = 0$). This reflects the fact that AND and OR are linearly separable functions and thus are exactly described by a perceptron, which (as discussed above) is defined purely by direct dependencies.\cite{Muroga-01} For comparison, consider the XOR function, the classic example of a higher-order dependence that relies irreducibly on the specific combination of inputs $x_1$ and $x_2$ (Fig.~\ref{fig_maxEnt}g). We find that direct dependencies provide no information about the function ($S_\text{dir} = S_\text{tot}$), such that all the variability arises from either higher-order dependencies or latent stochasticity. Together, these results demonstrate how stochastic functions can be decomposed into their constituent parts. \\


\noindent {\myfont \large Identifying optimal inputs}


To study real neurons, we must first specify their inputs. In large-scale recordings, while synaptic connectivity is rarely known, we can infer the optimal inputs that provide the best description of a neuron's output. In a population of $N$ neurons, for a given output neuron, we would like to select the $n < N$ inputs that, once included in the maximum entropy model [Eq.~(\ref{eq_P})], reduce our uncertainty about the output $S_\text{dir}$ as much as possible (Methods). This minimax entropy problem is generally intractable;\cite{Lynn-15, Lynn-16} however, we provide an efficient algorithm that greedily identifies the locally optimal inputs at each step (Methods). The result is a set of $n$ inputs that, by minimizing the model uncertainty $S_\text{dir}$, also maximize the amount of variability captured by direct dependencies $S_\text{tot} - S_\text{dir}$.

\begin{figure}[t!]
\centering
\includegraphics[width = .9\textwidth]{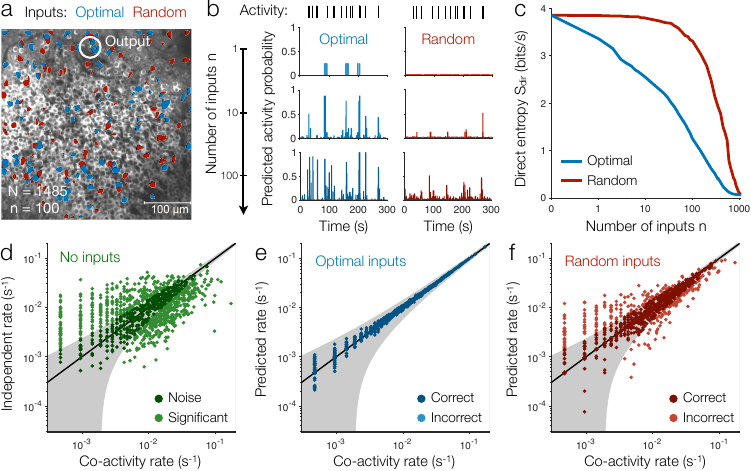} \\
\raggedright
\captionsetup{labelformat=empty}
{\spacing{1.25} \caption{\small \textbf{Fig.~\ref{fig_neuron} $|$ Identifying a minimal set of optimal inputs.} \textbf{a}, Population of $N = 1485$ neurons in the mouse hippocampus recorded as the animal runs along a virtual track (Methods).\cite{Gauthier-01, Meshulam-03} For a randomly-selected output neuron (circle), we illustrate the $n = 100$ optimal input neurons (blue) and an equal number of random inputs (red). \textbf{b}, Within a randomly-selected five-minute window, we plot the activity of the output neuron (top) and activity probabilities predicted by the maximum entropy model [Eq.~(\ref{eq_P})] for increasing numbers of optimal inputs (left) and random inputs (right). \textbf{c}, Direct entropy $S_\text{dir}$ versus the number of inputs $n$ for optimal inputs (blue) and random inputs (red). \textbf{d}, Co-activity rates between the output and all other neurons versus those predicted by the independent model with no inputs; line indicates equality, and shaded region indicates experimental errors (two standard deviations). \textbf{e}, With $n^* = 350$ optimal inputs, the model correctly predicts all remaining co-activity rates, and thus all direct dependencies on other neurons in the population (Methods). \textbf{f}, With the same number of random inputs, the model fails to predict many correlations. \label{fig_neuron}}}
\end{figure}

Consider a large population of neurons in the mouse hippocampus (Methods).\cite{Gauthier-01} These cells play key roles in encoding the animal’s location, mapping features in its environment, and storing memories of past events;\cite{Moser-01, Okeefe-01, Hafting-01, Meshulam-03} yet it remains unclear whether these functions arise from simple input-output dependencies. For a given output neuron, we infer the optimal $n$ inputs and the corresponding maximum entropy model (Fig.~\ref{fig_neuron}a). We only consider inputs that co-activate with the output at least once during a recording, which guarantees that the direct dependencies are well-defined (Methods). As the number of inputs increases, the minimal model quickly becomes expressive, making increasingly accurate predictions for the output activity (Fig.~\ref{fig_neuron}b). In fact, despite being maximally random with respect to interactions between inputs and time-delayed dependencies, the direct entropy $S_\text{dir}$ drops exponentially with the number of inputs (Fig.~\ref{fig_neuron}c). This means that with only a relatively small number of inputs, direct dependencies capture an exponentially large amount of variability $S_\text{tot} - S_\text{dir}$.

By contrast, with random inputs, the maximum entropy model fails dramatically, explaining effectively none of the variability until almost all the inputs are included (Fig.~\ref{fig_neuron}c). This suggests that by including all possible inputs, we risk overfitting the output, thus leading to an artificially low entropy $S_\text{dir}$. To avoid overfitting, we introduce the following regularization: we select the minimal number of inputs $n^*$ needed to predict all other direct dependencies in the population (Methods). This ensures that we do not fit any input-output dependencies that the model already predicts (Supplementary Information). The result is a combined framework for inferring the minimal model of direct dependencies [Eq.~(\ref{eq_P})] with the minimal set of inputs.


For the neuron in Fig.~\ref{fig_neuron}, among its correlations with other neurons, $66\%$ are significant, meaning that $34\%$ can be predicted with no inputs at all (Fig.~\ref{fig_neuron}d). We identify a minimal set of $n^* = 350$ inputs that are sufficient to predict all of the direct dependencies on other neurons (Fig.~\ref{fig_neuron}e). With only $n^*/(N-1) = 24\%$ of the possible inputs, these direct dependencies alone explain $(S_\text{tot} - S_\text{dir})/S_\text{tot} = 89\%$ of the neuron's variability. For comparison, with the same number of randomly-selected inputs, many of the correlations with other neurons remain unexplained (Fig.~\ref{fig_neuron}f). These findings establish that the vast majority of variability in activity, at least for one cell, is captured by a relatively small number of direct dependencies. \\

\noindent {\myfont \large Direct dependencies across systems and species}

\begin{figure}[t!]
\centering
\includegraphics[width = .9\textwidth]{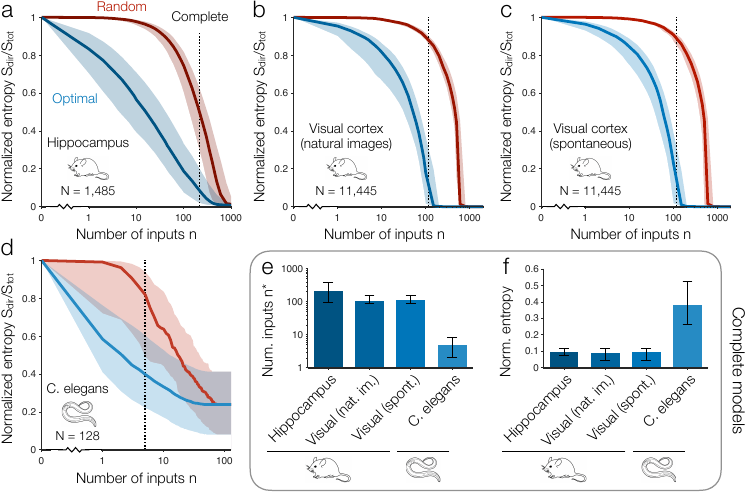} \\
\raggedright
\captionsetup{labelformat=empty}
{\spacing{1.25} \caption{\small \textbf{Fig.~\ref{fig_minimax} $|$ Small numbers of direct dependencies explain activity across systems and species.} \textbf{a}, Direct entropy $S_\text{dir}$ normalized by total entropy $S_\text{tot}$ for $n$ inputs chosen optimally (blue) or randomly (red). Lines and shaded regions represent medians and interquartile ranges across all $N = 1485$ hippocampal neurons in Fig.~\ref{fig_neuron}.\cite{Gauthier-01, Meshulam-03} Dashed line indicates the minimal number of inputs $n^*$ needed to capture all the direct dependencies for the median neuron; note that this number varies between neurons. \textbf{b-d}, Normalized model entropy $S_\text{dir}/S_\text{tot}$ versus number of inputs $n$ for 100 random output neurons within a population of $N = 11,445$ cells in the mouse visual cortex during responses to natural images (\textbf{b}) and spontaneous activity (\textbf{c}),\cite{Stringer-01} and for $N = 128$ neurons in the brain of \textit{C. elegans} (\textbf{d}).\cite{Dag-01} See Methods for experimental details. \textbf{e-f}, For the complete models in \textbf{a-d}, we compare the minimal number of inputs $n^*$ needed to capture all direct dependencies (\textbf{e}) and the normalized entropies $S_\text{dir}/S_\text{tot}$ (\textbf{f}). Values and error bars represent medians and interquartile ranges across neurons. \label{fig_minimax}}}
\end{figure}

We repeat this calculation for many neurons spanning different brain regions and species. In all cases, we study large recordings of spatially contiguous populations with recurrent connectivity, such that each neuron may receive synaptic inputs from the others. Across the $N=1485$ neurons in the hippocampal recording (Fig.~\ref{fig_neuron}a),\cite{Gauthier-01, Meshulam-03} we confirm that the direct entropy $S_\text{dir}$ drops exponentially with the number of inputs (Fig.~\ref{fig_minimax}a). This decrease is so sharp that, for the median neuron, the first input explains 17\% of the variability, and only 15 inputs are needed to explain 50\% of the variability. With $n^* = 214$ inputs (only $14\%$ of the population), the maximum entropy model correctly predicts the direct dependencies on all other neurons. These ``complete" models, which capture all of a neuron's direct dependencies, explain over 90\% of a neuron's entropy $S_\text{tot}$. This leaves less than 10\% of the variability for higher-order dependencies, time-delayed dependencies, and latent stochasticity combined.

In an even larger population of $N > 10^4$ cells in the mouse visual cortex,\cite{Stringer-01} for each neuron, one might expect that more inputs are needed to capture all of the direct dependencies. However, when responding to natural images, the median neuron only requires $n^* = 108$ inputs (less than 1\% of the entire population) to predict the remaining 99\% of its direct dependencies (Fig.~\ref{fig_minimax}b). Moreover, just as in the hippocampus, these complete models explain 91\% of each neuron's variability. For spontaneous activity in the same population, we observe nearly identical results (Fig.~\ref{fig_minimax}c). Thus, across the hippocampus and visual cortex, neurons are consistently described by input-output functions that (i) involve purely direct dependencies and (ii) are nearly deterministic; in other words, by perceptrons.



Finally, in the roundworm \textit{C. elegans}, one can record from the entire brain. This means that for each cell, we have access to nearly all its synaptic and extrasynaptic inputs.\cite{Dag-01, Randi-01} For the median neuron, a minimal set of $n^* = 5$ inputs is sufficient to predict all other direct dependencies in the brain. These remarkably simple computations explain 62\% of each neuron's variability $S_\text{tot}$ (Fig.~\ref{fig_minimax}d). Together, these findings (summarized in Fig.~\ref{fig_minimax}e-f) comprise our main result: that neuronal activity, spanning multiple systems and species, is explained by simple direct dependencies on only a small number of inputs. We confirm that these results hold for the average neuron (rather than median), are robust to downsampling the neural activity, and remain consistent across time (Supplementary Information).  \\

\noindent {\myfont \large Higher-order and time-delayed dependencies}


Complex functions require networks of artificial neurons.\cite{Hertz-01, Block-01, Muroga-01} In real neurons, however, interactions between dendrites and extrasynaptic signals can lead to higher-order dependencies that are responsible for gating and linearly non-separable functions like XOR (Fig.~\ref{fig_maxEnt}g).\cite{Beniaguev-01, Gidon-01, Losonczy-01, Polsky-01, Takahashi-01, London-01, Sykova-01, Poirazi-02, Park-01} Similarly, neurons can integrate inputs over time to execute important temporal computations and produce complex dynamics.\cite{Adelson-01, Shadlen-01, Loewenstein-02, Shi-01} However, the above results suggests that, with knowledge of only the direct, equal-time dependencies on individual inputs, one should be able to predict the higher-order dependencies on combinations of inputs as well as the time-delayed dependencies on past inputs. If true, this would paint a surprisingly simple picture in which higher-order and time-delayed dependencies arise naturally from direct, instantaneous dependencies.


To explain the $k^\text{th}$-order dependence $P(y|x_1,\hdots, x_k)$, it is sufficient to predict the correlations between the output $y$ and all subsets of the $k$ inputs (Methods). For each 2${}^\text{nd}$-order dependence $P(y|x_i,x_j)$, because our complete models capture all of the direct dependencies (either by fitting or prediction), all that remains is the triplet correlation between $y$, $x_i$ and $x_j$. In the hippocampus, the complete models predict 99.85\% of the triplet correlations (within experimental errors), leaving only 0.15\% of the 2${}^\text{nd}$-order dependencies unexplained by simpler direct dependencies (Fig.~\ref{fig_corr}a). For comparison, with the same numbers of random inputs, direct dependencies fail to explain many of the 2${}^\text{nd}$-order dependencies (Fig.~\ref{fig_corr}a, \textit{inset}). Returning to optimal inputs, the accuracy of the complete models increases as we study dependencies of even higher order. Direct dependencies fail to predict only 0.11\% of the quadruplet correlations (Fig.~\ref{fig_corr}b), and this fraction drops to 0.08\% for quintuplet correlations (Fig.~\ref{fig_corr}c). These results in the mouse hippocampus are recapitulated in the mouse cortex and \textit{C. elegans} (Fig.~\ref{fig_corr}d-g). We therefore find that the vast majority of higher-order dependencies can be understood as arising from simple direct dependencies, without relying on interactions between inputs.

\begin{figure}
\centering
\includegraphics[width = .9\textwidth]{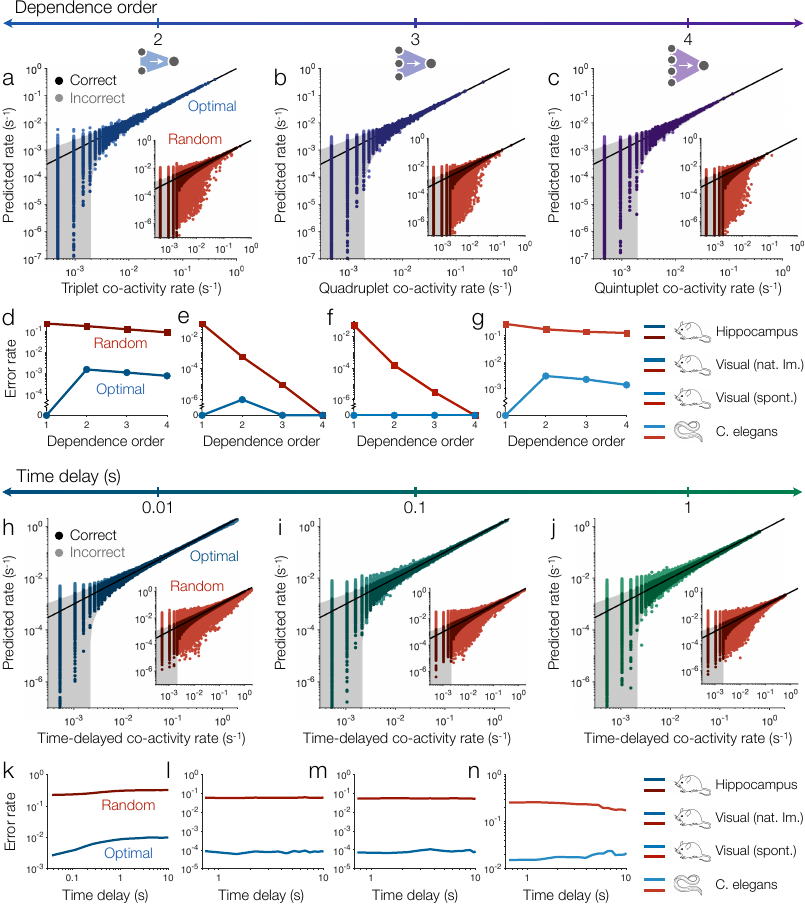} \\
\raggedright
\captionsetup{labelformat=empty}
{\spacing{1.25} \caption{\small \textbf{Fig.~\ref{fig_corr} $|$ Predicting higher-order and time-delayed correlations.} \textbf{a}, For the hippocampal population,\cite{Gauthier-01, Meshulam-03} we plot triplet co-activity rates predicted by the complete models versus those measured in data. Line indicates equality, shaded region indicates experimental errors (two standard deviations), dark points are correct (within errors), and light points are incorrect. Inset shows the co-activity rates predicted by maximum entropy models with the same numbers of random inputs. \textbf{b-c}, Quadruplet (\textbf{c}) and quintuplet (\textbf{d}) co-activity rates predicted by models with optimal inputs and random inputs (\textit{inset}). In \textbf{a-c}, for each output neuron we consider 100 correlations with randomly-selected groups of neurons. \textbf{d-g}, Fractions of correlations not predicted by direct dependencies in the mouse hippocampus (\textbf{d}),\cite{Gauthier-01} the mouse visual cortex during responses \label{fig_corr}}}
\end{figure}
\addtocounter{figure}{-1}
\begin{figure}[t!]
\centering
\raggedright
\captionsetup{labelformat=empty}
{\spacing{1.25} \caption{\small  to natural images (\textbf{e}) and spontaneous activity (\textbf{f}),\cite{Stringer-01} and the brain of \textit{C. elegans} (\textbf{g}).\cite{Dag-01} \textbf{h-j}, Comparison of time-delayed co-activity rates predicted by the complete models versus measured in data for delays of $0.1\,$s (\textbf{h}), $1\,$s (\textbf{i}), and $10\,$s (\textbf{j}) in the mouse hippocampus. Insets show predictions of maximum entropy models with the same numbers of random inputs. \textbf{k-n}, Fractions of time-delayed correlations not predicted by direct dependencies in the hippocampus (\textbf{k}),\cite{Gauthier-01} visual cortex responding to natural images (\textbf{l}) and during spontaneous activity (\textbf{m}), and \textit{C. elegans} (\textbf{n}). See Supplementary Information for details.}}
\end{figure}

Thus far, we have focused on the instantaneous dependencies between neurons within the same window of time (Fig.~\ref{fig_maxEnt}a). Yet the activity of each neuron $y$ at time $t$ may depend on the states of other neurons $x_i$ at previous times $t' < t$. To predict these time-delayed dependencies $P(y(t)|x_i(t'))$ using our complete models, it is sufficient to predict the time-delayed correlations between $y$ and $x_i$ (Methods). In the hippocampal recording, time is discretized into windows of length $\Delta t = 0.03\,$s.\cite{Gauthier-01, Meshulam-03} Despite being maximally random with regard to correlations longer than $\Delta t$, the complete models still predict 99.6\% of the correlations with time delay $0.1\,$s (Fig.~\ref{fig_corr}h), 99.1\% with delay $1\,$s (Fig.~\ref{fig_corr}i), and 99.0\% with delay $10\,$s (Fig.~\ref{fig_corr}j). We observe similarly high accuracy in the visual cortex and \textit{C. elegans} (Fig.~\ref{fig_corr}k-n). By contrast, with random inputs, direct dependencies fail to predict orders of magnitude more of the time-delayed correlations (Fig.~\ref{fig_corr}k-n). Thus, we find that most time-delayed dependencies are explained by simple instantaneous dependencies, with no information about the neural dynamics. \\

\noindent {\myfont \large Inferred neural network and robustness}


\begin{figure}[t!]
\centering
\includegraphics[width = \textwidth]{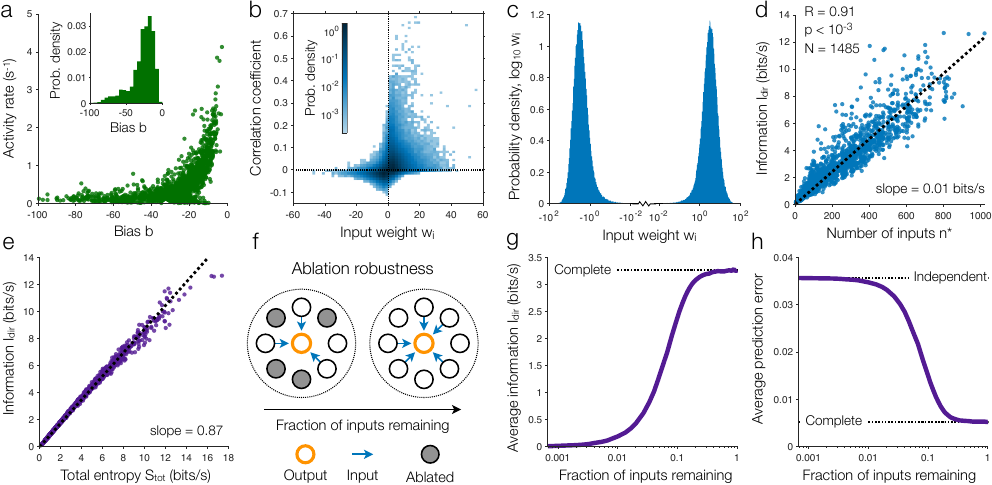} \\
\raggedright
\captionsetup{labelformat=empty}
{\spacing{1.25} \caption{\small \textbf{Fig.~\ref{fig_error} $|$ Model structure and network robustness.} \textbf{a}, Firing rate versus inferred bias $b$ for each neuron in the hippocampal population and distribution of inferred biases (\textit{inset}).\cite{Gauthier-01} \textbf{b}, Probability density of the correlation coefficient and corresponding input weight $w_i$ over all input-output pairs. \textbf{c}, Distribution of inferred weights $w_i$ over all input-output pairs in the hippocampal population. \textbf{d}, Direct information $I_\text{dir}$ versus number of inputs $n^*$ across all neurons; dashed line indicates linear fit. \textbf{e}, Information $I_\text{dir}$ versus total entropy $S_\text{tot}$ across all neurons; dashed line indicates linear fit. \textbf{f}, Illustration of robustness analysis. We remove (or ablate) neurons from the population by marginalizing over their states and study the predicted activity for the remaining neurons (Methods). \textbf{g}, Information $I_\text{dir}$ as a function of the fraction of inputs remaining for each neuron; dashed line illustrates the value for the full models with all inputs. \textbf{h}, Prediction error for the complete models with different fractions of the inputs removed; dashed lines indicate the values for the original models with all inputs (bottom) and independent models with no inputs (top). In panels \textbf{g} and \textbf{h}, values are averaged over neurons and 100 repeats of simulated ablations (Methods). \label{fig_error}}}
\end{figure}

Given their equivalence with logistic artificial neurons, we can study the structure of the inferred maximum entropy models in the context of neural computation (Fig.~\ref{fig_maxEnt}c). In the hippocampal population, all cells have negative biases $b$, leading them to favor silence over activity (Fig.~\ref{fig_error}a). Meanwhile, because the weights $w_i$ induce correlations between model neurons (Fig.~\ref{fig_error}b), it is tempting to interpret them as synaptic connection strengths. Fundamentally, however, these weights are defined to match the direct dependencies $P(y|x_i)$ measured in data, which may arise from extrasynaptic signals or shared dependencies on latent variables (such as unobserved neurons).\cite{Sykova-01, Morrell-01, Dag-01, Randi-01, Gauthier-01, Meshulam-03} Yet despite profound difficulties in deriving connectivity from activity,\cite{Das-01} we find that the inferred weights $w_i$ exhibit four key features of synaptic connections. First, as discussed above, the weights are sparse, with only a small number of inputs $n^*$ needed to explain all of a neuron's direct dependencies (Fig.~\ref{fig_minimax}e). Second, the distribution of magnitudes is heavy-tailed (specifically log-normal), with some rare weights that are orders of magnitude stronger than average (Fig.~\ref{fig_error}c). Third, the weights are evenly split between positive and negative, suggesting a delicate balance between excitatory and inhibitory interactions (Fig.~\ref{fig_error}c). Finally, unlike most existing maximum entropy models,\cite{Schneidman-01, Meshulam-03, Lynn-15} the weights are highly directed, with the weight from neuron $i$ to neuron $j$ differing significantly from its inverse. These sparse, heavy-tailed, balanced, and directed weights are universal features of synaptic connectivity observed across brain regions and species.\cite{Lin-02, Loomba-01, Lynn-13, Liu-01, VanVreeswijk-01, Lynn-05}

The connections between neurons enable the flow of information; for each neuron, the mutual information between inputs and output is equal to the drop in entropy $I_\text{true} = S_\text{tot} - S_\text{true}$.\cite{Strong-01, Cover-01} While this information is impossible to estimate directly from data, the maximum entropy models provide a tractable lower bound, equal to the amount of variability explained by direct dependencies $I_\text{dir} = S_\text{tot} - S_\text{dir} \le I_\text{true}$ (Methods). Moreover, since direct dependencies capture nearly all of the variability in activity (Fig.~\ref{fig_minimax}f), we know that this lower bound is tight, with $0.9I_\text{true} \lesssim I_\text{dir} \le I_\text{true}$ for the mouse hippocampus and visual cortex. Across neurons, we find that this direct information increases linearly with the number of inputs $n^*$, with each input communicating $0.01\,$bits/s to the output on average (Fig.~\ref{fig_error}d).

By symmetry, $I_\text{dir}$ also defines the amount of information that each neuron encodes about the rest of the population. For each bit generated by a neuron, we find that a consistent $0.87\,$bits encode information about its inputs (Fig.~\ref{fig_error}e). This large proportion of information concentrated on a small number of inputs indicates a highly redundant neural code. As in a Hopfield network, specifying the states of a small number of cells should be sufficient to predict the rest.\cite{Hopfield-01, Amit-01, Schneidman-01} To test this hypothesis, we can artificially remove, or ablate, some cells within a population by marginalizing over their states (Methods). For each of the remaining neurons, we then investigate the impact on the complete model (Fig.~\ref{fig_error}f). As each neuron loses more of its inputs, the flow of information from inputs to output undergoes a sharp transition (Fig.~\ref{fig_error}g). Above this transition, neurons can lose nearly 90\% of their inputs without impacting the flow of information, while below the transition, almost no information is communicated. Similarly, we can remove most of the inputs to a neuron before our model fails to accurately predict its activity (Fig.~\ref{fig_error}h). These findings demonstrate that the inferred neural network is strikingly robust, with each neuron maintaining nearly the same output activity even after losing the vast majority of its inputs. \\





\noindent {\myfont \large Discussion}

Despite intricate morphologies and biophysical dynamics,\cite{Beniaguev-01, Gidon-01, Losonczy-01, Polsky-01, Takahashi-01, London-01, Jan-01, Poirazi-01, Petersen-01, Poirazi-02, Park-01} neurons have long been studied using models of simple dependencies.\cite{Mcculloch-01, Hopfield-01, Rosenblatt-01, Hertz-01, Block-01, Amit-01} Here, we develop a framework to study whether neuronal activity arises from the simplest possible dependencies: those that capture the responses to individual inputs, but contain no information about interactions between inputs. Across the mouse hippocampus and visual cortex,\cite{Gauthier-01, Meshulam-03, Stringer-01} these direct dependencies explain over 90\% of the variability in neuronal activity (Fig.~\ref{fig_minimax}), leaving only 10\% for interactions between inputs, time-delayed dependencies, and latent variables (Fig.~\ref{fig_maxEnt}). Moreover, the inferred models---which are equivalent to artificial neurons---predict the higher-order dependencies on combinations of inputs and the time-delayed dependencies on past inputs (Fig.~\ref{fig_corr}) and recover salient features of synaptic connectivity (Fig.~\ref{fig_error}).

These results raise future questions about the nature of dependencies between neurons. As experiments advance to record from larger populations across species, neural systems, and imaging modalities,\cite{Stringer-01, Gauthier-01, Dag-01, Demas-01, Urai-01, Steinmetz-02} does neuronal activity consistently arise from direct dependencies? Of particular interest are electrophysiological recordings, which have sufficient temporal resolution to resolve individual spikes (Supplementary Information).\cite{Buzsaki-02, Wei-01} However, current large-scale recordings (for example, using Neuropixels\cite{Steinmetz-02,Steinmetz-03}) probe spatially elongated or discontiguous populations, potentially limiting the study of direct dependencies between neurons. Additionally, while our results suggest that most time-delayed dependencies are explained by instantaneous dependencies (Fig.~\ref{fig_corr}), one can immediately generalize our framework to include those that are not (Supplementary Information).\cite{Pillow-01} What do these significant time-delayed dependencies reveal about neural dynamics? Finally, as discussed above, the inferred neural network reflects not only causal interactions, but also functional correlations due to latent variables.\cite{Sykova-01, Morrell-01, Dag-01, Randi-01, Gauthier-01, Meshulam-03, Das-01} If the underlying population is defined by an Ising model---equivalent to a stochastic Hopfield network\cite{Hopfield-01} or Boltzmann machine\cite{Ackley-01}---we show that the inferred weights recover the true underlying interactions (Supplementary Information). As experiments mapping the wiring between neurons continue to advance,\cite{White-01, Lin-02, Loomba-01, Dorkenwald-01, Shapson-01} how does the inferred functional connectivity relate to underlying synaptic connectivity? The framework presented here provides the tools to begin answering these questions.


\newpage

\noindent {\large \myfont \textbf{Methods}}
\vspace{-28pt}

\noindent\rule{\textwidth}{.5pt}

\begin{methods}

\subsection{Maximum entropy model}. Consider a binary output $y \in \{0,1\}$ and a set of $n$ binary inputs $\bm{x} = \{x_1,\hdots, x_n\} \in \{0,1\}^n$. From experiments, we have $L$ samples of activity $y(\ell)$ and $\bm{x}(\ell)$, where $\ell = 1,\hdots,L$. From this data, we can estimate the direct dependencies $P(y|x_i)$ for all inputs $i=1,\hdots,n$. We want to derive the model $P(y|\bm{x})$ that is consistent with these direct dependencies and has maximum entropy
\begin{equation}
\label{eq_S}
S(P) = -\Big< \sum_y P(y|\bm{x})\log P(y|\bm{x})\Big>_{\bm{x}},
\end{equation}
where $\langle f(\bm{x}) \rangle_{\bm{x}} = \frac{1}{L}\sum_\ell f(\bm{x}(\ell))$ denotes an empirical average over the inputs, and (unless otherwise specified) we use log base two such that entropy is measured in bits.\cite{Jaynes-01, Cover-01} Each direct dependence $P(y|x_i)$ is uniquely defined by the averages $\langle y\rangle = \frac{1}{L}\sum_\ell y(\ell)$ and $\langle x_i\rangle = \frac{1}{L}\sum_\ell x_i(\ell)$ and the pairwise correlation $\langle yx_i\rangle = \frac{1}{L}\sum_\ell y(\ell)x_i(\ell)$. For every model $P(y|\bm{x})$, we have $\langle x_i\rangle_P = \langle x_i\rangle$, where $\langle f(y,\bm{x}) \rangle_P = \langle \sum_y f(y,\bm{x}) P(y|\bm{x})\rangle_{\bm{x}}$ denotes a model average. Thus, one only needs to constrain the average firing rate $\langle y\rangle$ and the correlations $\langle yx_i\rangle$ for all inputs. This maximum entropy model is known to take the logistic form,
\begin{equation}
\label{eq_P2}
P(y|\bm{x}) = \frac{1}{Z(\bm{x})}e^{y\big(b + \sum_i w_ix_i\big)},
\end{equation}
where $Z(\bm{x}) = 1 + e^{b + \sum_i w_ix_i}$ ensures normalization.\cite{Berger-01, Huang-01} The bias $b$ and weights $w_i$ are Lagrange multipliers that force the conrstraints $\langle y\rangle_P = \langle y\rangle$ and $\langle yx_i\rangle_P = \langle yx_i\rangle$. As discussed above, these maximum entropy models are equivalent to logistic models, which are a special case of generalized linear models that have provided key insights into neural dynamics.\cite{Pillow-01, Goris-01, Stevenson-01, Weber-01, Ostojic-01, Mensi-01, Churchland-01, Priebe-01}

\subsection{Computing model parameters.} Even with the functional form for the model, one must still compute the bias $b$ and weights $w_i$ so that the model matches the experimental average $\langle y\rangle$ and correlations $\langle yx_i\rangle$ for all inputs $i$. To do so, we minimize the Kullback-Leibler (KL) divergence $D_\text{KL}(Q||P)$ between the model and the empirical distribution $Q(y|\bm{x})$, which is equivalent to maximum likelihood estimation.\cite{Cover-01} Specifically, we perform gradient descent in the KL divergence, with gradients given by $\nabla_b D_\text{KL}(Q||P) = \langle y\rangle_P - \langle y\rangle$ and $\nabla_{w_i} D_\text{KL}(Q||P) = \langle yx_i\rangle_P - \langle yx_i\rangle$. This algorithm converges efficiently, even for very large $n$.

\subsection{Information in direct dependencies.} The true entropy of the neuron $S_\text{true}$ defines the latent variability that cannot be explained by dependencies on the inputs; however, unless the number of inputs $n$ is small, $S_\text{true}$ cannot be estimated directly from data. The total variability of the neuron with no knowledge of the inputs is defined by the entropy $S_\text{tot}$ of the marginal $P(y)$.\cite{Strong-01} Between these two extremes, with knowledge of only the direct dependencies $P(y | x_i)$, the entropy (in nats) of the maximum entropy model [Eq.~(\ref{eq_P2})] is given by
\begin{equation}
S_\text{dir} = \langle \log Z(\bm{x}) \rangle_{\bm{x}} - b\langle y\rangle - \sum_i w_i \langle yx_i\rangle.
\end{equation}
These entropies form a hierarchy $S_\text{tot} \ge S_\text{dir} \ge S_\text{true} \ge 0$. The difference $I_\text{true} = S_\text{tot} - S_\text{true}$ is the true mutual information between the inputs and the output.\cite{Cover-01} The difference $I_\text{dir} = S_\text{tot} - S_\text{dir}$, which lower-bounds $I_\text{true}$, is the mutual information between inputs and output in the model [Eq.~(\ref{eq_P})]. Finally, due to the maximum entropy form of the model, the KL divergence with the true firing probabilities $P_\text{true}(y|\bm{x})$ also simplifies to a difference in entropies $D_\text{KL}(P_\text{true}||P) = S_\text{dir} - S_\text{true}$.\cite{Cover-01} Thus, in the limit that the model entropy $S_\text{dir}$ becomes small, we know that $D_\text{KL}(P_\text{true}||P)$ also becomes small, and the model is exact. Moreover, in this limit the output becomes a deterministic function of the inputs $P(y=1|\bm{x}) = \Theta\big(b + \sum_iw_ix_i\big)$, where $\Theta(\cdot)$ is the step function. Together, these observations reveal that if $S_\text{dir}$ is small, then we have $S_\text{dir} \approx S_\text{true} \approx 0$, and the neuron itself becomes equivalent to an MP neuron or perceptron.\cite{Mcculloch-01, Block-01, Hertz-01}

\subsection{Optimal inputs.} For a given output neuron, we seek the $n$ inputs that produce the most accurate model [Eq.~(\ref{eq_P})]. As discussed above, the KL divergence between the model and the true firing probabilities reduces to a difference in entropies $D_\text{KL}(P_\text{true}||P) = S_\text{dir} - S_\text{true}$. Thus, the optimal inputs, which give the most accurate predictions for the output, are the ones that produce the maximum entropy model $P(y|\bm{x})$ with minimum entropy $S_\text{dir}$. This is an instance of the minimax entropy principle, which provides a general strategy for selecting optimal constraints in maximum entropy models.\cite{Lynn-15, Lynn-16, Carcamo-01}

\subsection{Greedy algorithm.} Searching for the optimal $n$ inputs among the $N-1$ possibilities is generally infeasible. Instead, we propose a greedy algorithm for growing a locally optimal set of inputs. We begin with the independent model $P(y)$, which has no inputs. We then fit a different model $P(y|x_i)$ for each of the $N-1$ possible inputs; among these, the optimal input is the one that produces the model with minimum entropy $S_\text{dir}$. Repeating this process, we greedily select the optimal input (which minimizes the entropy $S_\text{dir}$) at each step until we reach the desired number of inputs $n$.

\subsection{Approximate change in entropy.} The above algorithm involves fitting $O(nN)$ separate models: one for each of the $O(N)$ possible new inputs during each of the $n$ steps. To improve efficiency, rather than fitting a different model for each possible input, we can approximate the drop in entropy $\Delta S_\text{dir}$ analytically. Using perturbation theory, for a candidate neuron $i$ we expand the change in entropy in the limit of small prediction errors $\langle yx_i\rangle - \langle yx_i\rangle_P$, yielding an analytic approximation for $\Delta S_\text{dir}$ (Supplementary Information). Using this approximation to select the optimal input at each step, the greedy algorithm only requires fitting $O(n)$ models.

\subsection{Neural data.} Our framework can be used to investigate any binarized recordings of neuronal activity across different neural systems and species. Since we are interested in understanding the mapping from inputs to output, we focus on large recordings of spatially contiguous populations, where, for each neuron, we may have access to some or most of its inputs. Such recordings are made possible by calcium imaging, wherein animals are genetically modified so that their neurons fluoresce in response to changes in calcium concentration, which in turn follows the electrical activity of the cells. This fluorescence is recorded using an optical microscope with sample period $\Delta t$. To study activity on the fastest available timescale, we use the sample period $\Delta t$ to binarize each neuron into active ($x_i = 1$) or silent ($x_i = 0$; Fig.~\ref{fig_maxEnt}a).

We study four recordings of neuronal activity, each measured in previous experiments: one in the mouse hippocampus, two in the mouse visual cortex, and one in the brain of the roundworm \textit{C. elegans}. In the hippocampus, we study $N = 1485$ neurons in the CA1 region as the mouse runs along a virtual track (Fig.~\ref{fig_neuron}a); activity is recorded with scanning period $\Delta t = 1/30$s.\cite{Gauthier-01} In the visual cortex, we study $N = 11,445$ neurons recorded with scanning period $\Delta t = 2/3$s as the mouse is exposed to two separate visual stimuli: natural images (Fig.~\ref{fig_minimax}b) or a grey screen to measure spontaneous activity (Fig.~\ref{fig_minimax}c).\cite{Stringer-01} In \textit{C. elegans}, we study $N = 128$ neurons comprising nearly the entire brain recorded as the animal moves freely with period $\Delta t = 1/1.7$s (Fig.~\ref{fig_minimax}d).\cite{Dag-01} In the hippocampus and \textit{C. elegans}, we construct models for all neurons; while in the visual cortex, we study 100 randomly-selected output neurons.

Due to the sizes of the populations, some neurons never co-fire during the length of a given recording, leading to vanishing correlations $\langle yx_i\rangle = 0$. To avoid overfitting and divergences in the model parameters, for each output neuron $y$, we only consider inputs $x_i$ that co-fired with the output at least once, thus yielding positive correlations $\langle yx_i\rangle > 0$.

\subsection{Minimal set of inputs.} Given an output $y$ and a desired number of inputs $n$, the greedy algorithm identifies the locally optimal set of inputs $\bm{x} = \{x_1,\hdots,x_n\}$; however, we still need a principled method for choosing $n$. At each stage of the greedy algorithm, we have a model $P(y|\bm{x})$ with $n$ inputs. We use this model to predict the correlations $\langle yx_i\rangle_P$ with the other $N- n - 1$ neurons in the population. If all of these predictions are correct---that is, if they match the true correlations $\langle yx_i\rangle$ within experimental errors---then including another input amounts to fitting statistical noise. Thus, for each output neuron, we continue selecting inputs greedily until we reach a number $n^*$ for which the model predicts all other correlations. Specifically, we terminate the greedy algorithm when $|\langle yx_i\rangle - \langle yx_i\rangle_P| \le 2\sqrt{\langle yx_i\rangle}$ for all neurons $i$ with positive correlations $\langle yx_i\rangle > 0$, where $\sqrt{\langle yx_i\rangle}$ is the standard deviation of $yx_i$ in the data (assuming Poisson statistics). In this way, $n^*$ defines the minimal number of inputs needed for the model to match all of the (positive) direct dependencies, either by fitting or prediction. We confirm that this avoids overfitting (Supplementary Information). 

\subsection{Higher-order and time-delayed dependencies.} The statistical dependencies between inputs and output are encoded in correlations. As discussed above, the direct dependence $P(y|x_i)$ is uniquely defined by the averages $\langle y\rangle$ and $\langle x_i\rangle$ and the correlation $\langle yx_i\rangle$. Similarly, assuming stationarity the time-delayed dependence $P(y(t)|x_i(t'))$, where $t' < t$, is defined by $\langle y\rangle$, $\langle x_i\rangle$, and the time-delayed correlation $\langle y(t)x_i(t')\rangle$. The 2${}^\text{nd}$-order dependence $P(y|x_i,x_j)$ is defined by the direct dependencies $P(y|x_i)$ and $P(y|x_j)$ plus the triplet correlation $\langle yx_ix_j\rangle$. Thus, given a model that matches the direct dependencies, predicting the 2${}^\text{nd}$-order dependencies amounts to predicting the triplet correlations, as in Fig.~\ref{fig_corr}a. More generally, given a model that matches all of the $(k-1)^\text{th}$-order dependencies, predicting $k^\text{th}$-order dependencies amounts to predicting the corresponding $(k+1)^\text{th}$-order correlations (Fig.~\ref{fig_corr}b,c).

\subsection{Ablation robustness}. To study the robustness of the inferred models, we artificially remove (or ablate) inputs and study how this impacts the predicted output (Fig.~\ref{fig_error}f).\cite{Meyes-01} Given a model $P(y|\bm{x})$, we remove an input $i$ by marginalizing over its activity, yielding a new model
\begin{equation}
\tilde{P}(y|\bm{x}) = \frac{\sum_{x_i}P(y|\bm{x})Q(\bm{x})}{\sum_{x_i} Q(\bm{x})},
\end{equation}
where $Q(\bm{x})$ is the empirical distribution over inputs. Note that we do not re-fit the model parameters, we simply marginalize the original model over the ablated inputs. After removing a given fraction of inputs, we compute the mutual information $S_\text{tot} - S(\tilde{P})$ between the output and the remaining inputs (Fig.~\ref{fig_error}g) as well as the prediction error $\frac{1}{L}\sum_\ell (1 - \tilde{P}(y(\ell)|\bm{x}(\ell)))$ (Fig.~\ref{fig_error}h). In practice, we marginalize over a specified fraction of the population and repeat the above analysis for each remaining neuron as the output. We then average over all of the output neurons and 100 random realizations of this marginalization process.

\end{methods}

\section*{Data Availability}

The data analyzed in this paper are openly available at \\ \texttt{github.com/ChrisWLynn/Minimal{\_}computation}.

\section*{Code Availability}

The code used to perform the analyses in this paper is openly available at \\ \texttt{github.com/ChrisWLynn/Minimal{\_}computation}.





\begin{addendum}

\item[Supplementary Information.] Supplementary text and figures accompany this paper.

\item[Acknowledgements.] We thank P.~Dixit, B.~Machta, D.~Clark, T.~Geiller, M.~Leighton, F.~Mignacco, D.~Carcamo, N.~Weaver, and Q.~Yu for enlightening discussions and comments on earlier versions of the paper. We also acknowledge support from the National Institutes of Health (NIH/NIGMS R35GM160188) and the Department of Physics, Quantitative Biology Institute, and Wu Tsai Institute at Yale University.
 
\item[Author Contributions.] C.W.L. conceived the project, designed the models, performed the analysis, and wrote the manuscript and Supplementary Information.
 
\item[Competing Interests.] The author declares no competing financial interests.
 
\item[Corresponding Author.] Correspondence and requests for materials should be addressed to C.W.L. \\(christopher.lynn@yale.edu).
 
\end{addendum}

\newpage

\noindent {\large \myfont \textbf{Supplementary Information}}
\vspace{-28pt}

\noindent\rule{\textwidth}{.5pt}

\noindent {\myfont \large 1.~Introduction}

In this Supplementary Information, we provide extended analysis and discussion to support the results in the main text. The sections are ordered based on their references in the main text. In Sec.~2, we derive an analytic approximation to the drop in entropy that results from including an additional input in the maximum entropy model; we use this approximation to speed up the greedy algorithm that identifies optimal inputs. In Sec.~3, we demonstrate that our procedure for selecting the complete number of inputs $n^*$ avoids overfitting the activity of the output neuron. In Sec.~4, we show that the central results in the main text hold for the average neuron, not just median. In Sec.~5, we show that the central results hold after downsampling the recordings in time. In Sec.~6, we show that the central results hold across different halves of the recordings. In Sec.~7, we show that direct dependencies predict the complex higher-order dependencies between neurons consistently across different species and neural systems. In Sec.~8, we explore the structures of the inferred weights in different neuronal populations. In Sec.~9, we study direct dependencies in an electrophysiological recording of neuronal spiking in the salamander retina. In Sec.~10, we demonstrate how the maximum entropy model can be generalized to include time-delayed dependencies. Finally in Sec.~11, we demonstrate that the maximum entropy model achieves exact inference in the Ising model, which has long been used as a simplified model of recurrent neuronal activity. \\

\noindent {\myfont \large 2.~Approximate change in entropy}

Consider a binary output $y \in \{0,1\}$ and $n-1$ binary inputs $\bm{x} = \{x_1,\hdots, x_{n-1}\} \in \{0,1\}^{n-1}$. The maximum entropy model consistent with the direct dependencies $P(y|x_i)$, where $i\in\{1,\hdots,n-1\}$, is given by
\begin{equation}
\label{eq_P}
P(y|\bm{x}) = \frac{1}{Z(\bm{x})} e^{y\big(b + \sum_{i = 1}^{n-1} w_ix_i\big)},
\end{equation}
where $Z(\bm{x}) = 1 + e^{b + \sum_{i = 1}^{n-1} w_ix_i}$ ensures normalization. For ease of derivation, we define $x_0 = 1$ and $w_0 = b$, yielding
\begin{equation}
P(y|\bm{x}) = \frac{1}{Z(\bm{x})} e^{y\sum_{i = 0}^{n-1} w_ix_i}.
\end{equation}
The entropy of the model takes the form
\begin{equation}
S_\text{dir} = \langle \log Z(\bm{x})\rangle_{\bm{x}} - \sum_{i = 0}^{n-1} w_i \langle yx_i\rangle,
\end{equation}
where $\langle \cdot\rangle_{\bm{x}}$ represents an empirical average over the inputs $\bm{x}$, and $\langle \cdot \rangle$ represents an empirical average over the inputs and the output $y$.

We seek an analytic approximation for the drop in entropy $\Delta S_\text{dir}$ after including a new input $x_n$ in the model. Using perturbation theory, we can expand $\Delta S_\text{dir}$ in the limit of a small prediction error $\langle yx_n\rangle - \langle yx_n\rangle_P$, where $\langle yx_n\rangle_P = \langle \sum_y y x_n P(y|\bm{x})\rangle_{\bm{x}}$ is the correlation predicted by the existing model (without $x_n$ as an input, or, equivalently, with $w_n = 0$). To second order, we have
\begin{equation}
\Delta S_\text{dir} = (\langle yx_n\rangle - \langle yx_n\rangle_P)\frac{d S_\text{dir}}{d \langle yx_n\rangle_P}\Bigg|_{w_n = 0} + \frac{1}{2}(\langle yx_n\rangle - \langle yx_n\rangle_P)^2 \frac{d^2 S_\text{dir}}{d \langle yx_n\rangle_P^2}\Bigg|_{w_n = 0}.
\end{equation}
The first derivative takes the form
\begin{equation}
\label{eq_dS}
\frac{d S_\text{dir}}{d \langle yx_n\rangle_P} = \frac{d \langle \log Z(\bm{x})\rangle_{\bm{x}}}{d \langle yx_n\rangle_P} - w_n - \sum_{i = 0}^{n-1} \left(\langle yx_i\rangle \frac{d w_i}{d \langle yx_n\rangle_P} + w_i \frac{d \langle yx_i\rangle}{d \langle yx_n\rangle_P}\right).
\end{equation}
From the maximum entropy constraints, we know that $\frac{d \langle yx_i\rangle}{d \langle yx_n\rangle_P} = 0$. We also have
\begin{equation}
\frac{d \langle \log Z(\bm{x})\rangle_{\bm{x}}}{d \langle yx_n\rangle_P} = \sum_{i = 0}^{n-1} \frac{\partial \langle \log Z(\bm{x})\rangle_{\bm{x}}}{\partial w_i} \frac{d w_i}{d \langle yx_n\rangle_P} = \sum_{i = 0}^{n-1} \langle yx_i\rangle_P \frac{d w_i}{d \langle yx_n\rangle_P} = \sum_{i = 0}^{n-1} \langle yx_i\rangle \frac{d w_i}{d \langle yx_n\rangle_P}.
\end{equation}
Plugging into Eq.~(\ref{eq_dS}), we have
\begin{equation}
\frac{d S_\text{dir}}{d \langle yx_n\rangle_P} = - w_n.
\end{equation}
Thus, to first order, $\Delta S_\text{dir}$ vanishes.

The second derivative is given by
\begin{equation}
\label{eq_dS2}
\frac{d^2 S_\text{dir}}{d \langle yx_n\rangle_P^2} = -\frac{d w_n}{d \langle yx_n\rangle_P} = -\left(\frac{d \langle yx_n\rangle_P}{d w_n}\right)^{-1}.
\end{equation}
We have
\begin{equation}
\label{eq_dC}
\frac{d \langle yx_n\rangle_P}{d w_n} = \frac{\partial \langle yx_n\rangle_P}{\partial w_n} + \sum_{i = 0}^{n-1} \frac{\partial \langle yx_n\rangle_P}{\partial w_i} \frac{d w_i}{d w_n}.
\end{equation}
The derivative $\frac{d w_i}{d w_n}$ arises from the fact that, as $w_n$ changes, the existing weights $w_i$ must change to maintain the maximum entropy constraints $\langle yx_i\rangle_P = \langle yx_i\rangle$. Thus, we have
\begin{equation}
0 = \frac{d \langle yx_i\rangle_P}{d w_n} = \frac{\partial \langle y x_i\rangle_P}{\partial w_n} + \sum_{j = 0}^{n-1} \frac{\partial \langle y x_i\rangle_P}{\partial w_j} \frac{d w_j}{d w_n}.
\end{equation}
For $i,j \in \{0,\hdots,n-1\}$, we define the matrix
\begin{align}
M_{ij} &= \frac{\partial \langle yx_i\rangle_P}{\partial w_j} \\
&= \left< \frac{\partial}{\partial w_j} \frac{1}{Z(\bm{x})} \sum_y yx_i e^{y\sum_{k = 0}^{n-1} w_kx_k}\right>_{\bm{x}} \\
&= \left< \frac{1}{Z(\bm{x})}\sum_y y^2x_ix_j e^{y\sum_{k = 0}^{n-1} w_kx_k} - \frac{1}{Z(\bm{x})^2} \left(\sum_y yx_i e^{y\sum_{k = 0}^{n-1} w_kx_k} \right) \left(\sum_{y'} y'x_j e^{y'\sum_{k = 0}^{n-1} w_kx_k} \right) \right>_{\bm{x}} \\
&= \langle yx_ix_j\rangle_P - \langle yy'x_ix_j\rangle_P.
\end{align}
Similarly, for $i\in \{0,\hdots,n\}$, we have
\begin{equation}
 \frac{\partial \langle yx_i\rangle_P}{\partial w_n} = \langle yx_ix_n\rangle_P - \langle yy'x_ix_n\rangle_P.
\end{equation}
These definitions yield
\begin{equation}
\frac{d w_i}{d w_n} = -\sum_{j=0}^{n-1} (M^{-1})_{ij} \frac{\partial \langle yx_j\rangle_P}{\partial w_n}.
\end{equation}
Plugging into Eqs.~(\ref{eq_dS2}-\ref{eq_dC}), we have
\begin{equation}
\frac{d^2 S_\text{dir}}{d \langle yx_n\rangle_P^2} = -\left(\frac{d \langle yx_n\rangle_P}{d w_n}\right)^{-1}= - \left( \frac{\partial \langle yx_n\rangle_P}{\partial w_n} - \sum_{i,j = 0}^{n-1} \frac{\partial \langle yx_i\rangle_P}{\partial w_n} (M^{-1})_{ij}\frac{\partial \langle yx_j\rangle_P}{\partial w_n}\right)^{-1}
\end{equation}
We therefore arrive at an analytic approximation to the change in entropy
\begin{equation}
\Delta S_\text{dir} = -\frac{1}{2} \frac{(\langle yx_n\rangle - \langle yx_n\rangle_P)^2}{\frac{\partial \langle yx_n\rangle_P}{\partial w_n} - \sum_{i,j = 0}^{n-1} \frac{\partial \langle yx_i\rangle_P}{\partial w_n} (M^{-1})_{ij}\frac{\partial \langle yx_j\rangle_P}{\partial w_n}}.
\end{equation}
Note that this change is always negative, which follows from the fact that the maximum entropy must decrease with increasing constraints.

\noindent {\myfont \large 3.~Restricting inputs avoids overfitting}

When modeling the output of a neuron as a function of inputs using Eq.~(\ref{eq_P}), we need to make sure we are not overfitting the data, thus leading to an artificially low entropy $S_\text{dir}$. For a given output neuron, we use a greedy algorithm to iteratively select the $n$ optimal inputs that provide the best description of the output. Each time we add a new input $x_i$ to the model, we include an additional constraint on the direct dependence $P(y|x_i)$, which leads to a lower entropy $S_\text{dir}$. In large populations, such as those in the hippocampus and visual cortex studied in the main text, if we include all possible inputs to a given output neuron, then the entropy $S_\text{dir}$ drops to zero (Fig.~3 in the main text), and we have likely overfit the data.

To avoid overfitting, we terminate the greedy algorithm when the model is able to predict the direct dependencies $P(y|x_i)$ for all neurons $x_i$ that are not included as inputs in the model (within experimental errors). This ensures that we do not constrain any dependencies that the model already predicts, which would amount to fitting statistical noise. Note that we only consider neurons with positive correlations $\langle yx_i\rangle > 0$, such that the dependencies $P(y|x_i)$ are well-defined. Upon termination of the greedy algorithm, we arrive at a model (which we refer to as ``complete") with $n^*$ inputs that captures the direct dependencies $P(y|x_i)$ on all other neurons in the population (with $\langle yx_i\rangle > 0$), either by fitting or prediction.

Here, we demonstrate that by restricting the number of inputs we avoid overfitting. For each of the four recordings analyzed in the main text, we randomly divide the samples of activity into a training set (90\%) and test set (10\%). For each neuron, we infer the minimal computations in the training data as we increase the number of optimal inputs $n$. We then compute the negative log-likelihood of each model in both the training and test data,
\begin{equation}
\ell = -\langle \log P(y|\bm{x})\rangle_\text{train} \quad \quad \text{and} \quad \quad \ell_\text{test} = -\langle \log P(y|\bm{x})\rangle_\text{test},
\end{equation}
where $\langle\cdot\rangle_\text{train}$ and $\langle\cdot\rangle_\text{test}$ represent empirical averages over the training and test data, respectively. For the hippocampus and visual cortex, as we increase the number of inputs $n$, we see that the ratio $\ell_\text{test}/\ell$ remains small until we reach $n^*$; for $n > n^*$, the model begins fitting dependencies $P(y|x_i)$ that it can already predict, and $\ell_\text{test}$ increases dramatically relative to $\ell$ (Fig.~S\ref{fig_error}a-c). For \textit{C. elegans}, we see that the log-likelihood ratio $\ell_\text{test}/\ell$ doesn't increase until $n \gtrsim 10n^*$, which indicates that one might be able to increase the number of inputs $n^*$ in the complete model without overfitting (Fig.~S\ref{fig_error}d).

\addtocounter{figure}{-5}
\begin{figure}[t]
\centering
\includegraphics[width = \textwidth]{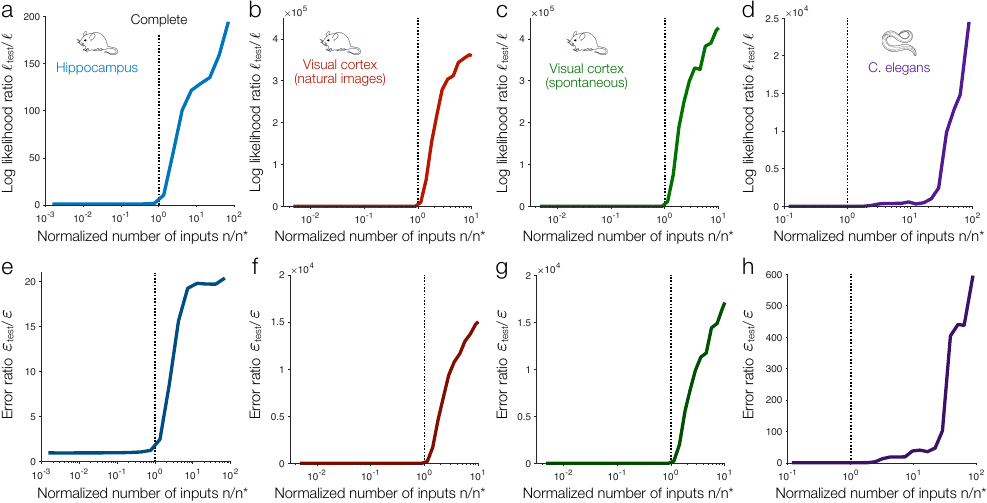} \\
\raggedright
\captionsetup{labelformat=empty}
{\spacing{1.25} \caption{\small \myfont \textbf{Fig.~S\ref{fig_error} $|$ Training and test errors.} \textbf{a-d}, Ratio of test and training log-likelihoods $\ell_\text{test}/\ell$ versus the number of inputs $n$ normalized by $n^*$ for populations of neurons in the mouse hippocampus (\textbf{a}),\cite{Gauthier-01} mouse visual cortex during responses to natural images (\textbf{b}) and spontaneous activity (\textbf{c}),\cite{Stringer-01} and the brain of \textit{C. elegans} (\textbf{d}).\cite{Dag-01} Dashed lines indicate the complete model with $n = n^*$. \textbf{e-h}, Ratio of test and training errors $\varepsilon_\text{test}/\varepsilon$ versus the number of inputs $n$ normalized by $n^*$ for the same data as \textbf{a-d}. Values in the hippocampus are averaged over all $N = 1485$ neurons in the population (\textbf{a} and \textbf{e}); values in the visual cortex are averaged over 100 randomly-selected neurons among a population of $N = 11,445$ (\textbf{b, c, f}, and \textbf{g}); and values in \textit{C. elegans} are averaged over all $N = 128$ neurons in the population (\textbf{d} and \textbf{h}). \label{fig_error}}}
\end{figure}

We also consider the model errors directly,
\begin{equation}
\varepsilon = \langle 1 - P(y|\bm{x})\rangle_\text{train} \quad \quad \text{and} \quad \quad \varepsilon_\text{test} = \langle 1 - P(y|\bm{x})\rangle_\text{test}.
\end{equation}
Just as for the log-likelihoods, across all populations, we find that the ratio $\varepsilon_\text{test}/\varepsilon$ remains small for $n \le n^*$ (Fig.~S\ref{fig_error}e-h). In the hippocampus and visual cortex, the test errors increase sharply for $n > n^*$, indicating overfitting (Fig.~S\ref{fig_error}e-g); while in \textit{C. elegans}, the test errors don't increase significantly until $n \gtrsim 10n^*$ (Fig.~S\ref{fig_error}h). Together, these results demonstrate that, by limiting the number of inputs to $n^*$, the complete models avoid overfitting the data. In turn, this tells us that the low entropies $S_\text{dir}$ of the complete models are not due to overfitting (Fig.~3f in the main text); instead, they indicate that the majority of neuronal variability is explained by direct dependencies on relatively small numbers of inputs.

\noindent {\myfont \large 4.~Average variability explained by direct dependencies}

\begin{figure}[t!]
\centering
\includegraphics[width = 0.9\textwidth]{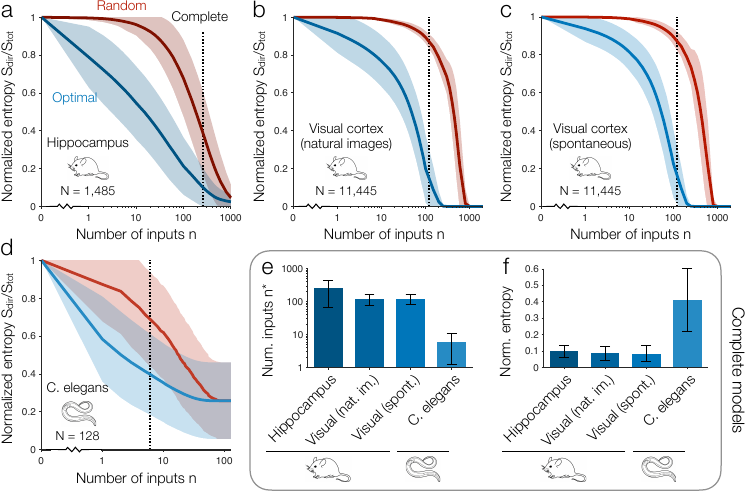} \\
\raggedright
\captionsetup{labelformat=empty}
{\spacing{1.25} \caption{\small \myfont \textbf{Fig.~S\ref{fig_mean} $|$ Average variability explained by direct dependencies.} \textbf{a}, Direct entropy $S_\text{dir}$ normalized by total entropy $S_\text{tot}$ for $n$ inputs chosen optimally (blue) or randomly (red). Lines and shaded regions represent means and one-standard-deviation error bars across all $N = 1485$ hippocampal neurons.\cite{Gauthier-01, Meshulam-03} Dashed line indicates the average minimal number of inputs $n^*$ needed to capture all the direct dependencies. \textbf{b-d}, Normalized model entropy $S_\text{dir}/S_\text{tot}$ versus number of inputs $n$ for 100 random output neurons within a population of $N = 11,445$ cells in the mouse visual cortex during responses to natural images (\textbf{b}) and spontaneous activity (\textbf{c}),\cite{Stringer-01} and for $N = 128$ neurons in the brain of \textit{C. elegans} (\textbf{d}).\cite{Dag-01} \textbf{e-f}, For the complete models in \textbf{a-d}, we compare the minimal number of inputs $n^*$ needed to capture all direct dependencies (\textbf{e}) and the normalized entropies $S_\text{dir}/S_\text{tot}$ (\textbf{f}). Values and error bars represent means and standard deviations across neurons. \label{fig_mean}}}
\end{figure}

In the main text, most quantities are provided as medians and interquartile ranges across neurons in each population. In Fig.~S\ref{fig_mean}, we show that the central results in the main text (Fig.~3) also hold for the average neuron. Specifically, if inputs are chosen optimally, we find that the direct entropy $S_\text{dir}$ drops sharply as a function of the number of inputs $n$ (Fig.~S\ref{fig_mean}a-d). Meanwhile, for the same numbers of random inputs, direct dependencies explain much less of the variability in activity. On average across the hippocampal neurons, we find that the complete models only require $n^* = 258$ inputs to capture all $N-1 = 1484$ of the direct dependencies in the population (Fig.~S\ref{fig_mean}e) and explain $1-S_\text{dir}/S_\text{tot} \approx 90\%$ of the variability in activity (Fig.~S\ref{fig_mean}f). On average across 100 cortical neurons, complete models only require $n^* = 121$ inputs to capture $N-1 = 11,444$ direct dependencies and explain $92\%$ of the total entropy; these numbers are consistent between spontaneous activity and responses to natural images. Finally, on average across the \textit{C. elegans} brain, complete models only require $n^* = 6$ inputs to capture $N-1 = 127$ direct dependencies and explain $59\%$ of the total entropy.

\noindent {\myfont \large 5.~Downsampling in time}

\begin{figure}[t]
\centering
\includegraphics[width = 0.9\textwidth]{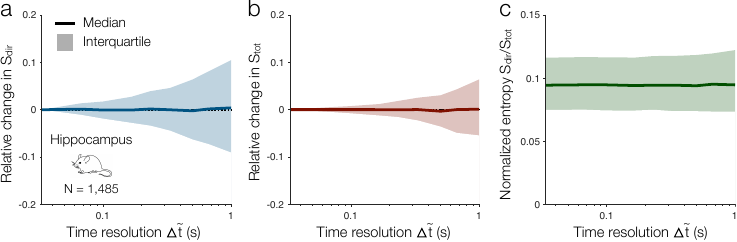} \\
\raggedright
\captionsetup{labelformat=empty}
{\spacing{1.25} \caption{\small \myfont \textbf{Fig.~S\ref{fig_down} $|$ Consistency under downsampling.} \textbf{a-c}, Relative change in direct entropy $S_\text{dir}$ (\textbf{a}), relative change in total entropy $S_\text{tot}$ (\textbf{b}), and change in normalized entropy $S_\text{dir}/S_\text{tot}$ (\textbf{c}) as functions of time resolution $\Delta \tilde{t}$ after downsampling. Lines and shaded regions represent medians and interquartile ranges across all $N=1485$ hippocampal neurons.\cite{Gauthier-01, Meshulam-03} Direct entropies $S_\text{dir}$ are computed using the complete maximum entropy models fit to the full data with empirical averages taken over subsamples of the input activity. \label{fig_down}}}
\end{figure}

For each of the recordings in the main text, we study activity on the fastest possible timescale, which is determined by the sample period $\Delta t$ of the experiment itself. One could instead study activity on longer timescales by downsampling the recording in time. To investigate the effect of downsampling, we consider the hippocampal population, which was recorded with the shortest period $\Delta t = 1/30\,$s. After randomly selecting one out of every $m$ samples, we compute the total entropy $S_\text{tot}$ and the entropy of the complete maximum entropy model $S_\text{dir}$ for each neuron with an effective time resolution of $\Delta \tilde{t} = m\Delta t$. As we increase the sample period $\Delta \tilde{t}$ up to one second, we find that $S_\text{tot}$, $S_\text{dir}$, and the normalized entropy $S_\text{dir}/S_\text{tot}$ all remain constant (Fig.~S\ref{fig_down}). This indicates that our results are robust to downsampling in time.

\noindent {\myfont \large 6.~Consistency over time}

\begin{figure}[t]
\centering
\includegraphics[width = 0.8\textwidth]{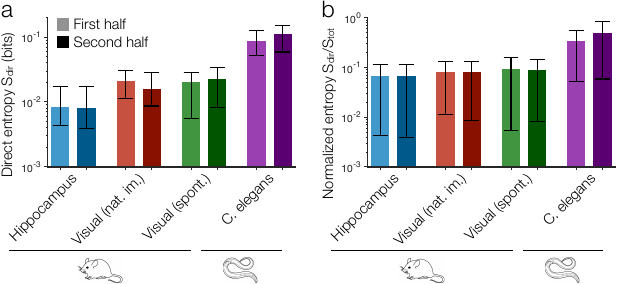} \\
\raggedright
\captionsetup{labelformat=empty}
{\spacing{1.25} \caption{\small \myfont \textbf{Fig.~S\ref{fig_halves} $|$ Consistency between halves of each recording.} \textbf{a-b}, Comparison of the direct entropies of the complete models $S_\text{dir}$ (\textbf{a}) and normalized entropies $S_\text{dir}/S_\text{tot}$ (\textbf{b}) between halves of each recording. Values and error bars represent medians and interquartile ranges across neurons. \label{fig_halves}}}
\end{figure}

In addition to the effects of downsampling, one can also study whether our results change over time. To do so, we divide each recording into halves and re-fit every model based on the statistics measured in each half of the data. We find that the direct entropy $S_\text{dir}$ of the complete models remains consistent between the first and second halves of each recording (Fig.~S\ref{fig_halves}a). Similarly, the normalized entropy $S_\text{dir}/S_\text{tot}$ (and therefore the fraction of entropy explained by direct dependencies $1 - S_\text{dir}/S_\text{tot}$) also remains consistent over time (Fig.~S\ref{fig_halves}b). This indicates that our central results do not change significantly over the length of each recording.

\noindent {\myfont \large 7.~Predicting higher-order dependencies}

Given the direct dependencies $P(y|x_i)$, the model in Eq.~(\ref{eq_P}) is maximally random with regard to all complex higher-order dependencies. Thus, if the model is capable of predicting the higher-order dependencies on two, three, or more inputs, then these can be viewed as emerging naturally from simpler direct dependencies. As discussed in the main text, predicting a given $k^\text{th}$-order dependence $P(y|x_1,\hdots,x_k)$ is equivalent to predicting the correlations between the output $y$ and all subsets of the $k$ inputs. Thus, if the model captures all of the $(k-1)^\text{th}$-order dependencies, all that remains is the $(k+1)^\text{th}$-order correlation $\langle yx_1\cdots x_k\rangle$.

\begin{figure}
\centering
\includegraphics[width = \textwidth]{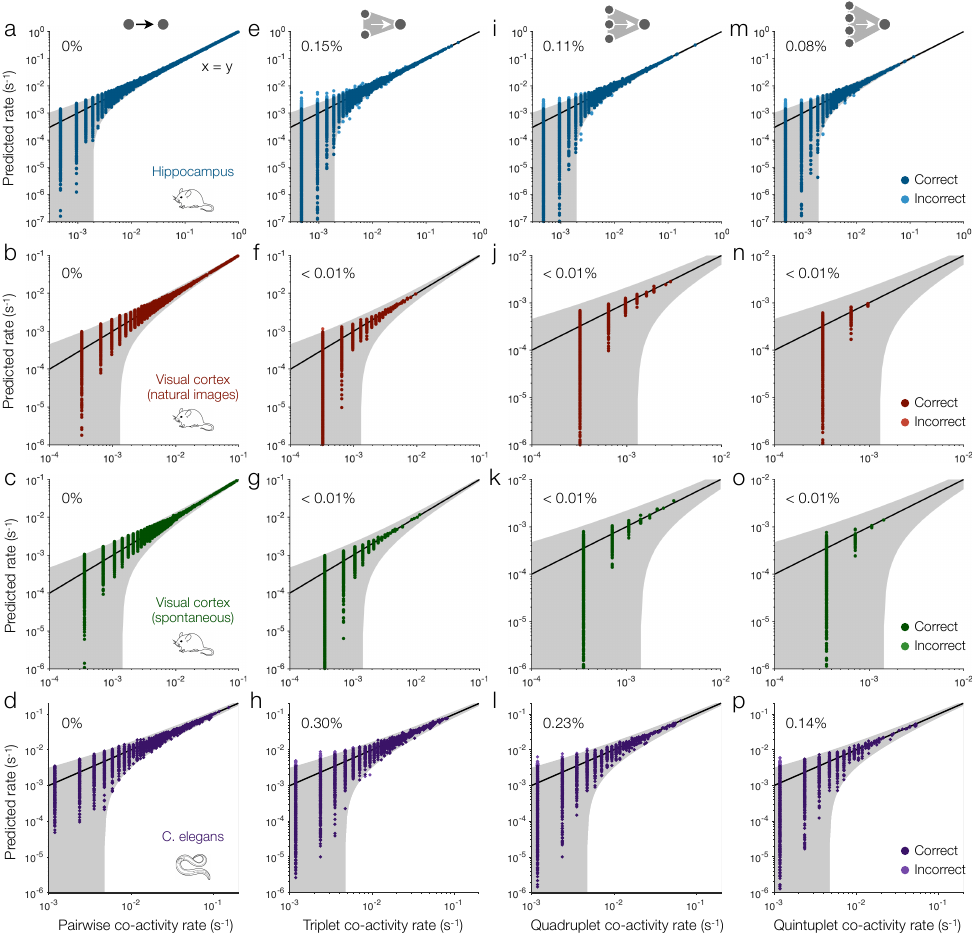} \\
\raggedright
\captionsetup{labelformat=empty}
{\spacing{1.25} \caption{\small \myfont \textbf{Fig.~S\ref{fig_corr} $|$ Predicting higher-order correlations.} \textbf{a-d}, Pairwise co-activity rates $\langle yx_i\rangle/\Delta t$ predicted by complete models versus those in the data for populations of neurons in the mouse hippocampus (\textbf{a}),\cite{Gauthier-01} mouse visual cortex during responses to natural images (\textbf{b}) and spontaneous activity (\textbf{c}),\cite{Stringer-01} and the brain of \textit{C. elegans} (\textbf{d}).\cite{Dag-01} Lines illustrate equality, shaded regions represent two standard deviations of experimental errors, dark points indicate correct predictions (within errors), light points indicate incorrect predictions, and percentages define the proportion of incorrect predictions. By the definition of the complete models, they correctly capture all pairwise co-activity rates, either by fitting or prediction. \textbf{e-p}, Triplet co-activity rates $\langle yx_ix_j\rangle/\Delta t$ (\textbf{e-h}), quadruplet co-activity rates $\langle yx_ix_jx_k\rangle/\Delta t$ (\textbf{i-l}), and quintuplet co-activity rates $\langle yx_ix_jx_kx_{\ell}\rangle/\Delta t$ (\textbf{m-p}) predicted by complete models versus those in the data for the \,\,\,\, \label{fig_corr}}}
\end{figure}
\addtocounter{figure}{-1}
\begin{figure}[t!]
\centering
\raggedright
\captionsetup{labelformat=empty}
{\spacing{1.25} \caption{\small \myfont same populations as \textbf{a-d}. For the mouse hippocampus (blue) and \textit{C. elegans} (purple) populations, we consider all $N = 1485$ and $N = 128$ neurons as outputs, respectively. For the mouse visual cortex (red and green), we consider 100 random output neurons within the population of $N = 11,445$ cells. For the pairwise co-activity rates (\textbf{a-d}), for each output neuron $y$ we consider all positive co-activities in the populations. For the higher-order co-activity rates (\textbf{e-p}), for each output neuron we consider 100 randomly-selected positive co-activities in the hippocampal (blue) and \textit{C. elegans} (purple) populations and $10^4$ randomly-selected positive co-activities in the visual cortex recordings (red and green).}}
\end{figure}

As discussed above, for each output neuron $y$, we use our greedy algorithm to identify the smallest number of inputs $n^*$ for which the model captures all of the positive correlations $\langle y x_i\rangle > 0$ in the population; this defines our complete model. Across each of the populations, we confirm that our complete models match all of the direct correlations within experimental errors (Fig.~S\ref{fig_corr}a-d) and, therefore, capture all of the direct dependencies $P(y|x_i)$.

Thus, to predict the second-order dependencies $P(y|x_i,x_j)$, we need only predict the triplet correlations $\langle yx_ix_j\rangle$. In Fig.~S\ref{fig_corr}i-l, we show that our models predict nearly all of the triplet correlations within experimental errors. Moreover, the minimal computations correctly predict an even larger proportion of the quadruplet (Fig.~S\ref{fig_corr}i-l) and quintuplet (Fig.~S\ref{fig_corr}m-p) correlations. Together, these results demonstrate that our minimal computation is capable of predicting the higher-order dependencies on combinations of two, three, or four inputs. In turn, this indicates that the vast majority of higher-order dependencies are explained simple direct dependencies, rather relying on complex interactions between inputs.

\noindent {\myfont \large 8.~Model structures}

\begin{figure}
\centering
\includegraphics[width = .85\textwidth]{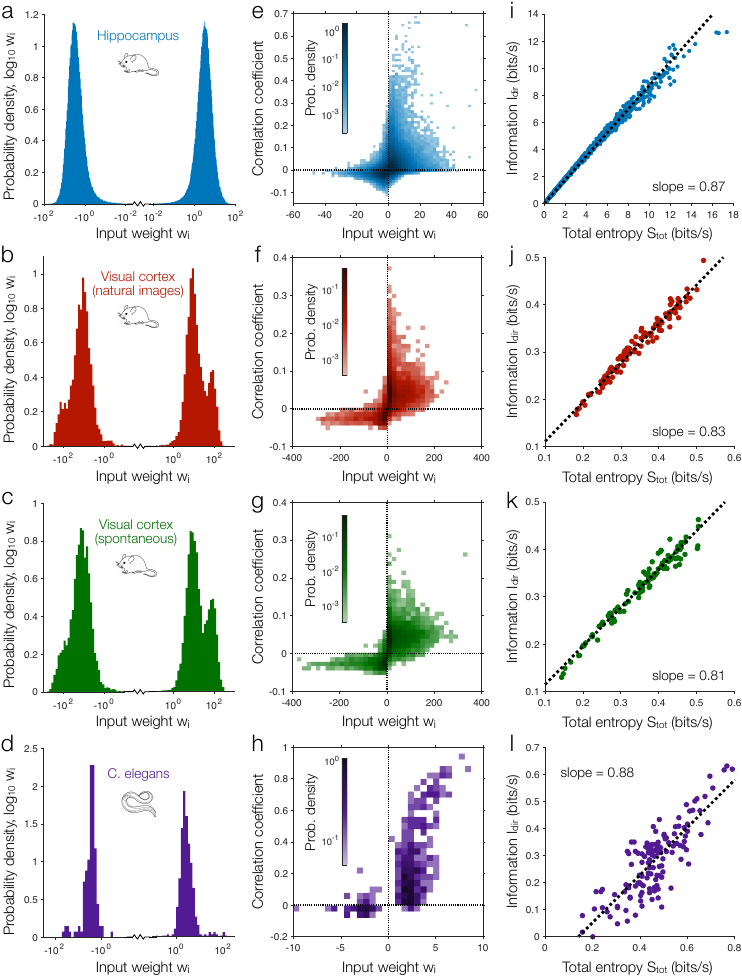} \\
\raggedright
\captionsetup{labelformat=empty}
{\spacing{1.25} \caption{\small \myfont \textbf{Fig.~S\ref{fig_model} $|$ Structure of complete models.} \textbf{a-d}, Distribution of inferred input weights $w_i$ over all complete models in populations in the mouse hippocampus (\textbf{a}),\cite{Gauthier-01} mouse visual cortex during responses to natural images (\textbf{b}) and spontaneous activity (\textbf{c}),\cite{Stringer-01} and the brain of \textit{C. elegans} (\textbf{d}).\cite{Dag-01} \textbf{e-h}, Probability density of the correlation coefficient and corresponding input weight $w_i$ over all input-output pairs in complete models for the same populations as \textbf{a-d}. \textbf{i-l}, Information $I_\text{dir}$ in complete models versus total entropy $\,\,\,\,\,\,\,\,\,\,\,\,\,\,\,\,\,\,\,\,\,\,\,\,\,\,\,\,\,$ \label{fig_model}}}
\end{figure}

After constructing the complete model for each output neuron, we have the opportunity to investigate the properties of the inferred weights. In the main text, we show in the mouse hippocampus that the inferred input weights $w_i$ exhibit four key features. First, the weights are sparse, with only a small number of inputs $n^*$ needed to explain all of a neuron's pairwise dependencies (Fig.~3e in the main text). Second, the distribution of magnitudes is heavy-tailed (specifically log-normal), with some rare weights that are orders of magnitude stronger than average (Fig.~S\ref{fig_model}a). Third, the weights are evenly split between positive and negative, suggesting a delicate balance between excitatory and inhibitory interactions (Fig.~S\ref{fig_model}a). Finally, unlike many existing maximum entropy models,\cite{Schneidman-01, Meshulam-03, Lynn-15} the weights are highly directed, with the connection strength from input $i$ to output $j$ differing significantly from the reverse. These sparse, heavy-tailed, balanced, and directed weights are universal features of synaptic connectivity observed across brain regions and species.\cite{Lin-02, Loomba-01, Lynn-13, Liu-01, VanVreeswijk-01, Lynn-05}

\addtocounter{figure}{-1}
\begin{figure}[t!]
\centering
\raggedright
\captionsetup{labelformat=empty}
{\spacing{1.25} \caption{\small \myfont $S_\text{tot}$ (normalized by $\Delta t$) across all output neurons; dashed line indicates linear fit. For the mouse hippocampus (blue) and \textit{C. elegans} (purple), we consider all $N = 1485$ and $N = 128$ neurons as outputs, respectively. For the mouse visual cortex (red and green), we consider 100 random output neurons within the population of $N = 11,445$ cells.}}
\end{figure}

For recordings in the mouse visual cortex and \textit{C. elegans}, we have already seen that the connectivity is sparse (Fig.~3e in the main text), with the complete models requiring an even smaller number of inputs $n^*$ than in the mouse hippocampus. In Fig.~S\ref{fig_model}b-d, we also confirm that the inferred weights $w_i$ are close to log-normally distributed (i.e., heavy-tailed) and quite evenly split between positive and negative (i.e., balanced). Since the weights are also directed, we find that the inferred connectivities share the same four key features across each of the recordings.

In the hippocampal population, we saw in the main text that positive (negative) input weights $w_i$ tend to produce positive (negative) correlations between the inputs and output in each model (Fig.~S\ref{fig_model}e). In Fig.~S\ref{fig_model}f-h, we observe the same relationship in the mouse visual cortex and \textit{C. elegans} recordings. Finally, we consider the mutual information $I_\text{dir} = S_\text{tot} - S_\text{dir}$ between the inputs and output in the complete models, which provides a tight upper bound on the true mutual information $I_\text{true} = S_\text{tot} - S_\text{true}$. In the hippocampus, we find that the mutual information increases linearly with the total entropy $S_\text{tot}$, with each bit generated by a neuron encoding a consistent 0.87 bits of information about its inputs (Fig.~S\ref{fig_model}i). We confirm similar linear relationships across the mouse visual cortex and \textit{C. elegans} recordings (Fig.~S\ref{fig_model}j-l). Together, these results suggest that the amount of information encoded in statistical dependencies may be strikingly consistent across species and neural systems.

\noindent {\myfont \large 9.~Electrophysiological activity}

By studying the statistical dependencies between neurons, we hope to capture aspects of the true underlying interactions. To have any hope of measuring the output of a neuron and many of its inputs simultaneously, we study recordings that meet three criteria. First, the recordings must be large. In \textit{C. elegans}, each neuron receives synaptic inputs from tens of others;\cite{White-01} in the mouse cortex, this number increases to thousands or tens of thousands.\cite{Loomba-01} We therefore focus on recordings that are at least this large. Second, we focus on neural systems with recurrent connectivity, such that each neuron within a recording may receive inputs from the others. Finally, we focus on populations that are spatially contiguous, which ensures that each pair of neurons is physically close and therefore has an opportunity to interact synaptically.

Current experiments that meet these criteria involve calcium imaging of large neuronal populations, as investigated in the main text.\cite{Gauthier-01, Stringer-01, Dag-01} Most electrophysiological recordings, by contrast, fail to meet at least one of the above criteria. For example, Neuropixel probes provide the largest electrophysiological recordings involving up to thousands of neurons.\cite{Steinmetz-02, Steinmetz-03} However, each probe records a spatially elongated one-dimensional slice through the brain, and large-scale recordings involve multiple probes distributed discontiguously throughout a single brain region or multiple regions.

\begin{figure}[t]
\centering
\includegraphics[width = 0.8\textwidth]{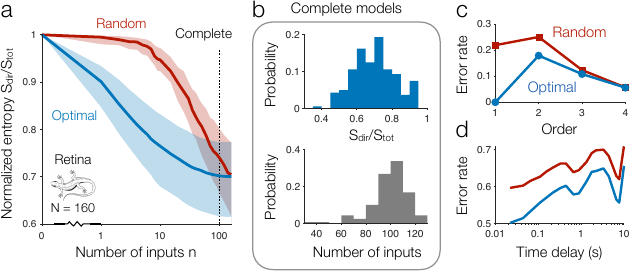} \\
\raggedright
\captionsetup{labelformat=empty}
{\spacing{1.25} \caption{\small \myfont \textbf{Fig.~S\ref{fig_ephys} $|$ Electrophysiological recording of retinal ganglion cells.} \textbf{a}, Direct entropy $S_\text{dir}$ normalized by total entropy $S_\text{tot}$ for $n$ inputs chosen optimally (blue) or randomly (red). Lines and shaded regions represent medians and interquartile ranges across all $N = 160$ neurons within an electrophysiological recording of ganglion cells in the salamander retina.\cite{Tkacik-02} Dashed line indicates the minimal number of inputs $n^*$ needed to capture all the direct dependencies for the median neuron. \textbf{b}, Distributions of normalized entropies $S_\text{dir}/S_\text{tot}$ (\textit{top}) and inputs $n^*$ (\textit{bottom}) for complete models over different retinal ganglion neurons. \textbf{c-d}, Fractions of higher-order dependencies and time-delayed dependencies not predicted by direct dependencies. \label{fig_ephys}}}
\end{figure}

Despite these experimental limitations, our framework can immediately be applied to study electrophysiological recordings. To demonstrate this generality, we investigate the activity of $N=160$ ganglion cells in the salamander retina recorded using a multi-electrode array in previous experiments.\cite{Tkacik-02} The electrical activity of each cell is binarized within time windows of width $\Delta t = 20\,$ms. For each neuron, we use our greedy algorithm to identify the optimal inputs. As the number of inputs $n$ increases, we find that the direct entropy $S_\text{dir}$ decreases exponentially before eventually plateauing at around 70\% of the total entropy $S_\text{tot}$ (Fig.~S\ref{fig_ephys}a). For random inputs, the direct entropy decreases much more slowly (Fig.~S\ref{fig_ephys}a). For the median neuron, one requires $n^* = 100$ inputs to explain all $N-1 = 159$ of the direct dependencies (Fig.~S\ref{fig_ephys}b, \textit{bottom}). These complete models explain $1 - S_\text{dir}/S_\text{tot} \approx 30\%$ of the total variability in activity (Fig.~S\ref{fig_ephys}b, \textit{top}). We note that this relatively low amount of variability explained by direct dependencies is not surprising given that retinal ganglion cells receive most of their information from photoreceptors. We therefore expect most of the variability in activity to arise from latent variables (Fig.~1d in the main text). Indeed, relative to the recordings analyzed in the main text, direct dependencies fail to predict a larger proportion of higher-order dependencies on multiple inputs (Fig.~S\ref{fig_ephys}c) as well as time-delayed dependencies on past inputs (Fig.~S\ref{fig_ephys}d). As electrophysiological recordings advance to record larger contiguous populations, our maximum entropy framework provides a foundation for studying direct dependencies in future work.

\noindent {\myfont \large 10.~Time-delayed dependencies}

In the main text, we demonstrate that the vast majority of time-delayed dependencies are explained by instantaneous direct dependencies (Fig. 4h-n in the main text). However, a small number of significant time-delayed dependencies remain unexplained. To illustrate how these time-delayed dependencies can be incorporated into our maximum entropy framework, we investigate a simulated population of $N=100$ neurons with known temporal interactions. For each neuron $i$, we draw time-delayed input weights $W_{ij}(\delta t)$ for every other neuron $j$ from a zero-mean Gaussian with standard deviation
\begin{equation}
\sigma(\delta t, \tau) = \frac{e^{-\delta t/\tau}}{\sum_{\delta t' = 0}^T e^{-\delta t'/\tau}},
\end{equation}
where $\delta t = 0,\hdots,T$ is the time-delay, $T$ is the length of the longest time-delayed interaction, and $\tau$ is a time constant that defines how quickly the interactions decay. We simulate the population dynamics based on the logistic model
\begin{equation}
\label{eq_dt}
P(x_i(t) =1\, |\, \bm{x}(t), \hdots, \bm{x}(t-T)) = \sigma \left[ \sum_{\delta t = 0}^T \sum_{j\neq i} W_{ij}(\delta t)x_j(t - \delta t) \right],
\end{equation}
where $\sigma(\cdot)$ is the sigmoid function and the vector $\bm{x}(t) = \{x_i(t)\}$ defines the activity of all neurons at time $t$.

\begin{figure}[t]
\centering
\includegraphics[width = 0.65\textwidth]{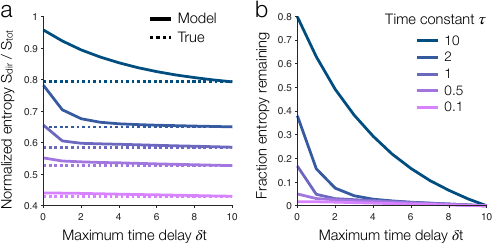} \\
\raggedright
\captionsetup{labelformat=empty}
{\spacing{1.25} \caption{\small \myfont \textbf{Fig.~S\ref{fig_dt} $|$ Maximum entropy models with time-delayed dependencies.} \textbf{a}, Normalized entropy $S_\text{dir}/S_\text{tot}$ for maximum entropy models constrained to match direct time-delayed dependencies $P(x_i(t)|x_j(t - \delta t))$ up to a maximum delay $\delta t$. Dashed lines indicate the normalized entropy $S_\text{true}/S_\text{tot}$ for the ground-truth models, which are defined in Eq.~(\ref{eq_dt}) with a maximum time delay $T = 10$. Colors reflect different time constants $\tau$. \textbf{b}, Fraction of available entropy not explained by maximum entropy models $(S_\text{dir} - S_\text{true})/(S_\text{tot} - S_\text{true})$ versus maximum time delay. In both panels, dynamics are simulated for $N = 100$ neurons and values are averages over all neurons. \label{fig_dt}}}
\end{figure}

Given simulated dynamics, we can construct the maximum entropy model consistent with not only instantaneous dependencies $P(x_i(t) | x_j(t))$, but also time-delayed dependencies $P(x_i(t) | x_j(t - \delta t))$. As we include more of the time-delayed dependencies in the maximum entropy model, the normalized entropy $S_\text{dir}/S_\text{tot}$ quickly plateaus to the true value $S_\text{true}/S_\text{tot}$ of the ground-truth model (Fig.~S\ref{fig_dt}a). As expected, this convergence is faster for dynamics in which the time-delayed interactions $w_{ij}(\delta t)$ decay more quickly (that is, for smaller time constants $\tau$; Fig.~S\ref{fig_dt}b). For $\tau = 1$, instantaneous dependencies explain $(S_\text{tot} - S_\text{dir})/(S_\text{tot} - S_\text{true}) \approx 83\%$ of the available variance, and this fraction increases to 98\% for $\tau = 0.1$. These results demonstrate that direct dependencies may be surprisingly effective in capturing time-delayed interactions.

\noindent {\myfont \large 11.~Exact inference in Ising model}

We have seen that the maximum entropy model in Eq.~(\ref{eq_P}) provides a tight approximation to the activity of real neurons. But how does the model perform in an artificial system where the underlying interactions are known? As a salient example, we consider Ising models, which are equivalent to Hopfield networks (with non-zero temperature),\cite{Hopfield-01} Boltzmann machines,\cite{Ackley-01} and maximum entropy models for joint activity.\cite{Schneidman-01, Meshulam-03, Lynn-15, Lynn-16} In each of these forms, the Ising model has provided key insights into neural computation, both in the brain and artificial networks. Here, we show that the maximum entropy model provides exact inference in the Ising model.

Consider a network with $N$ binary neurons defined by the state vector $\bm{x} = \{x_1,\hdots,x_N\}$. Each neuron has a bias $b_i$ that influences it toward activity or silence, and neurons interact via a symmetric interaction matrix $W_{ij} = W_{ji}$. The joint probability of a given activity state is defined by the Boltzmann distribution
\begin{equation}
P(\bm{x}) = \frac{1}{Z}e^{\sum_i b_ix_i + \frac{1}{2}\sum_{ij} W_{ij}x_ix_j},
\end{equation}
where
\begin{equation}
Z = \sum_{\bm{x} \in \{0,1\}^N} e^{\sum_i b_ix_i + \frac{1}{2}\sum_{ij} W_{ij}x_ix_j}
\end{equation}
is the partition function, which ensures normalization.

Suppose we want to study the output of one neuron (say $x_i$) in response to the remaining neurons $\bm{x}_{-i} = \{x_1,\hdots ,x_{i-1}, x_{i+1},\hdots ,x_N\}$. Specifically, we would like to compute the conditional probability $P(x_i|\bm{x}_{-i}) = P(\bm{x})/P(\bm{x}_{-i})$, where
\begin{equation}
P(\bm{x}_{-i}) = \sum_{x_i = 0,1} P(\bm{x}) = \frac{1}{Z} e^{\sum_{j\neq i} b_jx_j + \frac{1}{2} \sum_{jk\neq i} W_{ij}x_ix_j} \left(1 + e^{b_i + \sum_j W_{ij}x_j}\right).
\end{equation}
We therefore have
\begin{equation}
P(x_i|\bm{x}_{-i}) = \frac{e^{x_i(b_i + \sum_j W_{ij}x_j)}}{1 + e^{b_i + \sum_j W_{ij}x_j}} = \sigma \big(b_i + \sum_j W_{ij}x_j\big).
\end{equation}
This tells us that the conditional probability of one Ising variable in response to the rest of the system takes precisely the same form as the model in Eq.~(\ref{eq_P}). Thus, given the direct dependencies $P(x_i|x_j)$ for all $j\neq i$, the maximum entropy model is guaranteed to recover to the correct bias $b_i$ and weights $W_{ij}$. This approach provides an efficient method for inference in high-dimensional Ising models.\cite{Ravikumar-01}

\newpage

\section*{\large References}
\vspace{-30pt}
\noindent\rule{\textwidth}{.5pt}

\bibliographystyle{naturemag}
\bibliography{MaxEntBib}

\end{document}